\documentclass[aps,prb,notitlepage,superscriptaddress,amssymb,graphicx,11pt, floats]{revtex4-1}

\usepackage{graphicx}
\usepackage{etoolbox}
\usepackage{mathrsfs}
\usepackage{bm, upgreek}
\newcommand{\diff}{\mathrm{d}}
\usepackage{mathtools}
\def\be{\begin{equation}}
\def\ee{\end{equation}}

\patchcmd{\thebibliography}{\section*{\refname}}{}{}{}
\begin{document}
\title{Magnetostatic twists in room-temperature skyrmions explored by nitrogen-vacancy center spin texture reconstruction}

\author{Y.\ Dovzhenko$^{\dagger}$}
\affiliation{Department of Physics, Harvard University, 17 Oxford St., Cambridge, MA 02138, USA.}

\author{F.\ Casola$^{\dagger}$}
\affiliation{Department of Physics, Harvard University, 17 Oxford St., Cambridge, MA 02138, USA.}
\affiliation{Harvard-Smithsonian Center for Astrophysics, 60 Garden St., Cambridge, MA 02138, USA.}

\author{S.\ Schlotter}
\affiliation{John A. Paulson School of Engineering and Applied Sciences, Harvard University, Cambridge, MA 02138, USA.}
\affiliation{Department of Materials Science and Engineering, Massachusetts Institute of Technology, Cambridge, MA 02139, USA.}

\author{T.\ X.\ Zhou}
\affiliation{Department of Physics, Harvard University, 17 Oxford St., Cambridge, MA 02138, USA.}
\affiliation{John A. Paulson School of Engineering and Applied Sciences, Harvard University, Cambridge, MA 02138, USA.}

\author{F.~B\"uttner}
\affiliation{Department of Materials Science and Engineering, Massachusetts Institute of Technology, Cambridge, MA 02139, USA.}

\author{R.\ L.\ Walsworth}
\affiliation{Harvard-Smithsonian Center for Astrophysics, 60 Garden St., Cambridge, MA 02138, USA.}
\affiliation{Department of Physics, Harvard University, 17 Oxford St., Cambridge, MA 02138, USA.}

\author{G.\ S.\ D.\ Beach}
\affiliation{Department of Materials Science and Engineering, Massachusetts Institute of Technology, Cambridge, MA 02139, USA.}

\author{A.\ Yacoby$^{*,}$}
\affiliation{Department of Physics, Harvard University, 17 Oxford St., Cambridge, MA 02138, USA.}
\maketitle

\textit{$^{\dagger}$ These authors contributed equally to this work.}\\
\textit{$^*$ Correspondence and requests for materials should be addressed to this author.}\\

\textbf{Magnetic skyrmions are two-dimensional non-collinear spin textures characterised by an integer topological number\cite{Ref1,Ref2,Ref3}. Room-temperature skyrmions\cite{Ref13} were recently found in magnetic multilayer stacks, where their stability was largely attributed to the chiral Dzyaloshinskii-Moriya interaction (DMI)\cite{Ref6,Ref7} that arises due to the broken inversion symmetry at the interfaces \cite{Ref9,Ref10,Ref11,Ref12}. The strength of the DMI and its role in stabilizing the skyrmions, however, is not yet well understood, and imaging of the full spin structure is needed to address this question. Here, we use the single electron spin of a Nitrogen-Vacancy (NV) centre in diamond \cite{Ref15} to reconstruct an image of all three spin components of a skyrmion in a Pt/Co/Ta multilayer under ambient conditions. We introduce a new methodology to obtain, characterise, and unambiguously select physically meaningful solutions from the manifold of magnetization structures that produce the same measured stray field. We find that the skyrmion shows a N\'eel-type domain wall as expected, but the chirality of the wall is not left-handed, contrary to preceding reports of DMI in similar materials\cite{Belmeguenai,
Emori, Pizzini, Ryu}. Rather than being uniform through the film thickness as usually assumed, we propose skyrmion tube-like structures whose chirality rotates uniformly through the film thickness, due to a competition between the DMI and stray fields.
These results indicate 
that NV magnetometry, combined with our data reconstruction method, provides a unique tool to investigate this previously inaccessible phenomenon.}\\

Magnetic skyrmions are topological defects originally proposed as being responsible for the suppression of long-range order in the two-dimensional Heisenberg model\cite{Ref1,Ref2} at finite temperature. 
The earliest observations of magnetic skyrmions were reported in bulk crystals\cite{Ref4} of noncentrosymmetric ferromagnetic materials at cryogenic temperatures. 
Recently a new class of thin film materials has emerged, which support skyrmions at room temperature\cite{Ref9,Ref10,Ref11}. 
These results have paved the way towards spintronics applications and call for a quantitative and microscopic characterization of the novel spin textures. However, magnetic imaging of sputtered thin films at room temperature in the presence of variable external magnetic fields represents a serious experimental challenge for established techniques\cite{Ref11}, calling for a new approach.\\

We address this challenge using a magnetic sensor based on a single Nitrogen-Vacancy (NV) centre in diamond\cite{Ref15}. We record the projection on the NV axis of the magnetic field produced by the magnetization pattern in the film. This information is sufficient for reconstructing all three components of the magnetic field without the need for vector magnetometry\cite{Ref22} (see Section II of the Supplement). However, obtaining the underlying spin structure is an under-constrained problem\cite{Ref23}. System-dependent assumptions, e.g., regarding the spatial dependence of a certain spin component\cite{Ref24}, may artificially restrict the manifold of solutions compatible with experimental results. Here, we introduce a method to study such a manifold and show that we can classify all solutions by their helicity. We make use of an energetic argument to require continuity of the structure and discard unphysical solutions.\\

An overview of our scanning magnetometry setup is shown in Fig.~1a-c. The sample of interest is deposited on a quartz tip and scanned underneath a stationary diamond pillar, which contains a single NV centre about 30 nm below the surface. An image of a typical diamond pillar of approximately 200 nm diameter is shown in Fig.~1a. 
The sample consists of a sputtered [Pt (3 nm) / Co (1.1 nm) / Ta (4 nm)] x 10 stack with a seed layer of Ta (3 nm)\cite{Ref9}. We pattern 2 $\upmu$m diameter discs of this film on the flat surface of a cleaved quartz tip, pictured in Fig.~1c (see Methods and Section I of the Supplement). All measurements are performed in ambient conditions with a variable bias magnetic field delivered by a permanent magnet and aligned along the NV axis.\\
\begin{figure}[!]
\centering
\includegraphics[width=\columnwidth]{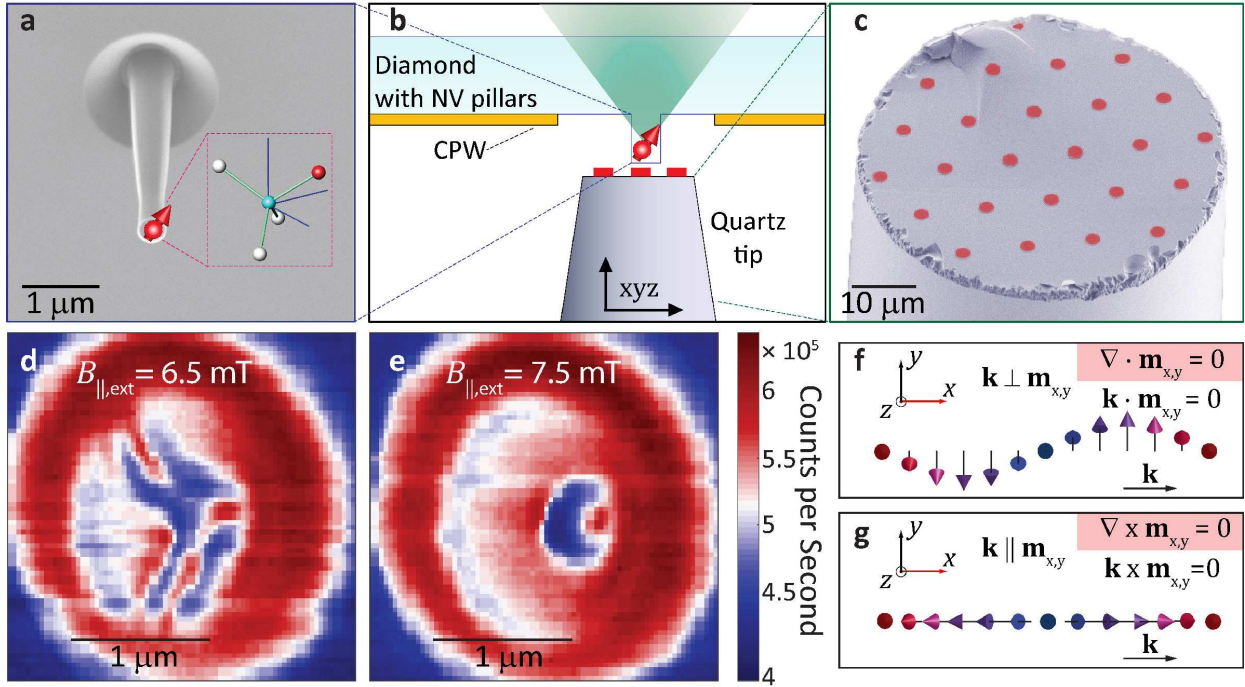}
\vspace{-0.5cm}
\caption{ \textbf{Experimental setup.} \textbf{a}: Electron microscope image of a typical diamond nanopillar containing a single NV centre approximately 30 nm deep. Rows of such pillars, $\sim$1.5 $\upmu$m tall, are located inside the gaps of a coplanar waveguide (CPW), which is evaporated on the surface of the diamond (see also \textbf{b}). The CPW is used to deliver the microwave excitations necessary to control the NV spin state. The inset shows schematically the geometry of an NV centre in a diamond lattice, pictured in greater detail in Fig.~2d. \textbf{b}: Sketch of the measurement configuration. A quartz tip with patterned magnetic discs is brought into contact with the diamond nanopillar. The quartz tip and the diamond are mounted on separate stacks of piezo-based positioners and scanners, enabling sub-nanometer movement along all the three $xyz$ axes. \textbf{c}: False-coloured electron microscopy image of a representative quartz tip, where 10 repetitions of a sputtered Pt(3nm)/Co(1.1nm)/Ta(4 nm) stack (red) are defined via electron beam lithography and subsequent lift-off as described in Section I of the Supplement. \textbf{d,e}: NV photoluminescence recorded at 6.5 mT (panel \textbf{d}) and 7.5~mT (panel \textbf{e}) external bias field. The optical excitation power is $\sim$100 $\upmu$W. Higher counts are observed above the magnetic disc due to reflection from the metallic surface. Within the disc boundary, areas with lower counts correspond to large stray magnetic fields perpendicular to the NV axis. \textbf{f}: Sketch of the Bloch-like spin configuration of a 1D magnetic spiral. Here, the local moments of the spiral rotate within a plane that forms an angle $\gamma$=$\pm \pi$/2 with respect to the propagation vector \textbf{k} of the magnetic structure (see text). \textbf{g}: Structure analogous to \textbf{f} for a N\'eel-like cycloid configuration. Here $\gamma$=0 ($\pi$) for spins rotating in the anticlockwise (clockwise) direction in the $zx$-plane.}
\label{fig:fig1}
\end{figure}
In order to identify magnetic features in the patterned discs, we employ a qualitative measurement scheme based on the rate of NV photoluminescence. In the presence of stray magnetic fields perpendicular to the NV axis, fewer red photons are emitted by the NV centre under continuous green excitation\cite{Ref25}. Two photoluminescence scans across the sample at different values of the bias magnetic field are shown in Fig.~1d,e. At 6.5~mT of external magnetic field, we observe a stripe-like modulation of the NV photoluminescence (see Fig.~1d). This pattern is reminiscent of the labyrinth domain arrangement of the local magnetization expected in these materials\cite{Ref5,Ref9}. When the bias field is increased by 1 mT the labyrinth domains collapse, forming a bubble-like feature shown in Fig.~1e. Our aim in the present paper is to determine the associated spin texture in this high-field regime.\\ 

To extract quantitative information, we use the NV magnetometer to measure two-dimensional (2D) spatial maps of the stray field component $B_{\parallel}$ parallel to the NV quantization axis (see Fig.~2a). The measurement plane $\boldsymbol{\uprho}=(x,y)$ is parallel to the magnetic film with the NV sensor at a distance d$\sim$30~nm from this surface. Since no free or displacement currents are present at the NV site, all information about the stray field $\mathbf{B}$ is contained in the magnetostatic potential $\phi_{\mathrm{M}}$, defined as $\mathbf{B}=-\nabla \phi_{\mathrm{M}}$. It follows that the three spatial components of $\mathbf{B}$ are linearly dependent in Fourier space, and all components of $\mathbf{B}$ at a distance $\geq d$ from the film can be obtained numerically from the map at $d$ using upward propagation\cite{Ref23} (see also Section IIB of the Supplement). These properties of magnetic fields allow us to reconstruct 2D maps for $B_{z}(\boldsymbol{\uprho},d)$ and $B_{x}(\boldsymbol{\uprho},d)$ (see Fig.~2b,c) from the 2D scan of $B_{\parallel}(\boldsymbol{\uprho},d)$. In these measurements, the bias field $\mathbf{B}_{\mathrm{ext}}$  is aligned with the quantization axis of the NV, which forms an angle $\mathrm{\theta}_{\mathrm{NV}} \approx 54.7^{\circ}$ with the axis $z$ normal to the magnetic film surface (see Fig.~2d). We independently confirm the component reconstruction procedure by comparing the reconstructed stray field magnitude perpendicular to the NV axis ($B_{\perp,\mathrm{r}}$ in Fig.~2e) to the one extracted from the experiment (Fig.~2f and Section IIA of the Supplement). The good agreement demonstrates our ability to perform vector magnetometry with only one NV orientation.\\

\begin{figure}[!]
\includegraphics[width=1\textwidth]{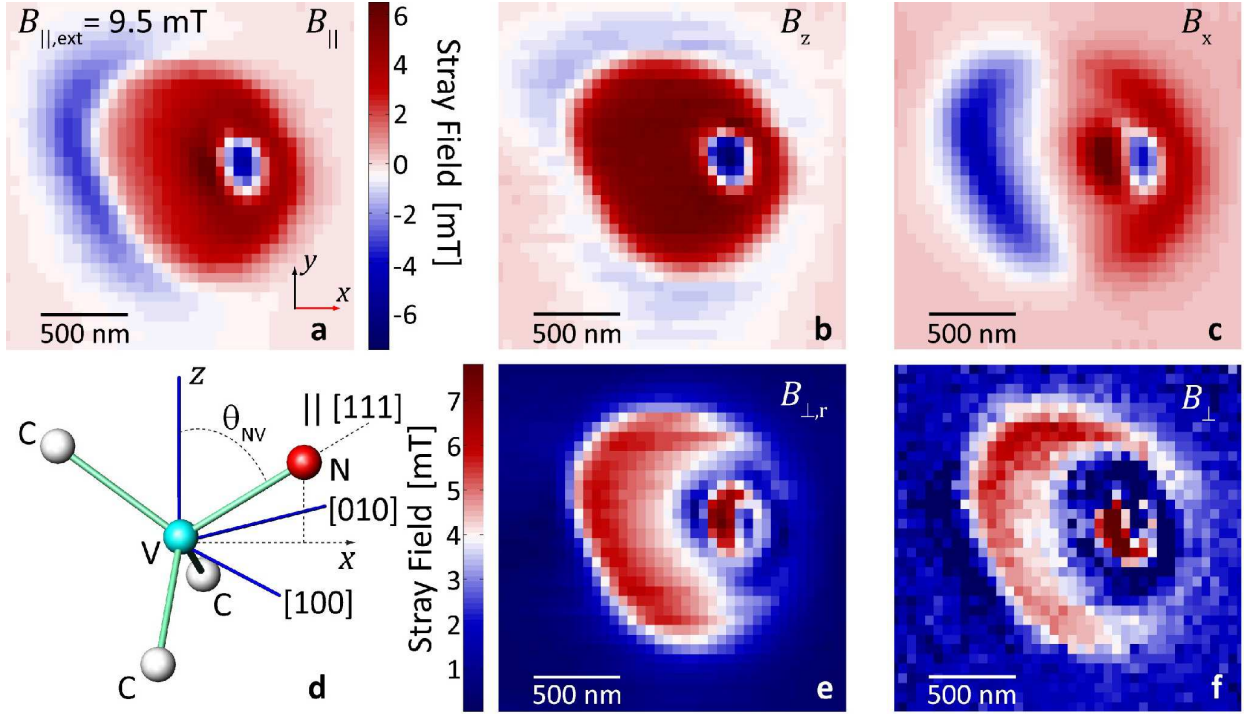}

\vspace{0.5cm}
\caption{ \textbf{Reconstruction of the magnetic stray field components. } \textbf{a}: 2D map of the stray field projection $B_{\parallel}$ on the NV axis (see also panel \textbf{d}). The measurement was performed at a bias field of  $B_{\parallel, \mathrm{ext}}=$9.5 mT applied along the [111] diamond axis. \textbf{b,c}: Reconstructed components of the stray field along the $z$ and $x$-directions, respectively. The $z$-direction is perpendicular to the magnetic disc. \textbf{d}: Sketch of the coordination geometry of a nitrogen-vacancy defect in diamond, illustrating the direction parallel to the quantization axis ($\parallel$) relative to the Cartesian reference frame of the setup $(x,z)$. Carbon, nitrogen, and vacancy sites are labeled C, N and V, respectively. The $z$-axis is orthogonal to the diamond surface. \textbf{e, f}: Reconstructed (\textbf{e}) and measured (\textbf{f}) magnitude of the stray field perpendicular to the NV centre [111] direction. The measured map is extracted from the spin level mixing of the NV (see Section IIA of the Supplement). The reconstructed plot is obtained using the procedure outlined in Section IIB of the Supplement.}
\label{fig:fig2}
\end{figure}

Because the components of $\mathbf{B}(\boldsymbol{\uprho},d)$  are not independent, they do not contain sufficient information for extracting the underlying spin structure. We will need additional criteria to narrow down the range of possible solutions (see Section IIB of the Supplement). We examine the out-of-plane field $B_{z}(\boldsymbol{\uprho},d)$, a component that fully preserves all the rotational symmetries of the out-of-plane magnetization. Starting with one magnetic layer and assuming that the local sample magnetization vector $\mathbf{m}(\boldsymbol{\uprho},z)=(\mathbf{m}_{x,y},m_{z})$ is the same throughout the layer thickness $t$, we show (see Section IIC of the Supplement and Ref.~\onlinecite{Ref26}) that $B_{z}(\boldsymbol{\uprho},d)$ has the following dependence on local magnetization:
\begin{equation}
B_{z}(\boldsymbol{\uprho},d) = - \frac{\upmu_0 M_s}{2} \left( \alpha_{z}(d,t) \ast \nabla^2 m_z(\bm{\uprho})  + \alpha_{x,y}(d,t) \ast \nabla \cdot \mathbf{m}_{x,y}(\bm{\uprho}) \right),
\label{eq:1}
\end{equation}
where $\ast$ denotes convolution in the $x,y$-plane, $M_s$ is the maximum value of the saturation magnetization in the disc, and we allow $0 \leq || \mathbf{m} || \leq 1$ to accommodate spatial dependence of the saturation magnetization of the film. Extension to multilayers is discussed in Section II and X of the supplement. The radially symmetric functions $\alpha_{z}(d,t)$ and $\alpha_{x,y}(d,t)$ are point spread functions, which account for the NV-to-film distance.\\

Since derivatives commute with convolutions, eq. \eqref{eq:1} is equivalent to Gauss's equation of the form $B_z= -\nabla \cdot \mathbf{F}$, where $B_z$ can be viewed as an effective local charge density and $\mathbf{F}$ as an effective electric field. The local magnetization components $\mathbf{m}_{x,y}$ and $m_z$ play the role of an effective vector and scalar potential, respectively. In analogy to standard electromagnetism\cite{Ref27}, potentials can be uniquely determined by fixing a gauge (see also Section III of the Supplement). Each gauge leads to a different spin helicity\cite{Ref5} $\gamma$ for the magnetic structure $\mathbf{m}$. For a simple helical structure, $\gamma$ is the angle between the plane of rotation of the local moments and the propagation vector\cite{Ref28}. For example, spirals (sketched in Fig.~1f) have helicity $\gamma$=$\pm \pi$/2 and are referred to as Bloch configurations in the context of domain walls\cite{Ref24}. The associated condition $\mathbf{k} \cdot \mathbf{m} = 0$ for this case can be also expressed as $\nabla \cdot \mathbf{m}_{x,y}=0$, resembling the Coulomb gauge in electromagnetism\cite{Ref27}. The opposite case is a spin cycloid (see Fig.~1g) with helicity $\gamma$=0 ($\pi$) representing a N\'eel-like arrangement of spins\cite{Ref24}. In this case $\nabla \times \mathbf{m}_{x,y}=0$. We show how to solve eq. \eqref{eq:1} for $\mathbf{m}$ in both Bloch and N\'eel gauges in Section III A,B and Section IV of the Supplement. This gauge approach allows us, for first time, to systematically identify the complete set of spin structures compatible with local magnetometry data.\\

For both gauges we use a numerical variational approach to find a spin structure whose stray field matches the measured field map. The measured field map is shown in Fig.~3a, while a simulated field map from a reconstructed spin structure is plotted in Fig.~3b. We plot cuts through the experimental map and the computed map along $x$ and $y$ axes in Fig.~3c. A 2D plot of the spin structure for the N\'eel (Bloch) gauge is shown in Fig.~3d (Fig.~3e). In our analysis we take into account local variations in the saturation magnetization by scaling the magnetization vector $\mathbf{m}$ to the $m_z$ value obtained in the saturated regime (see Fig.~3f and Section VII of the Supplement). The two structures in Fig.~3d,e are particular examples chosen from an infinite number of solutions to eq. \eqref{eq:1}. These solutions are stable with respect to variation in NV depth, as we demonstrate in section VIII of the Supplement, thus accounting for the inherent uncertainty of NV implantation depth estimation. \\
\begin{figure}[!]
\centering
\includegraphics[width=\textwidth]{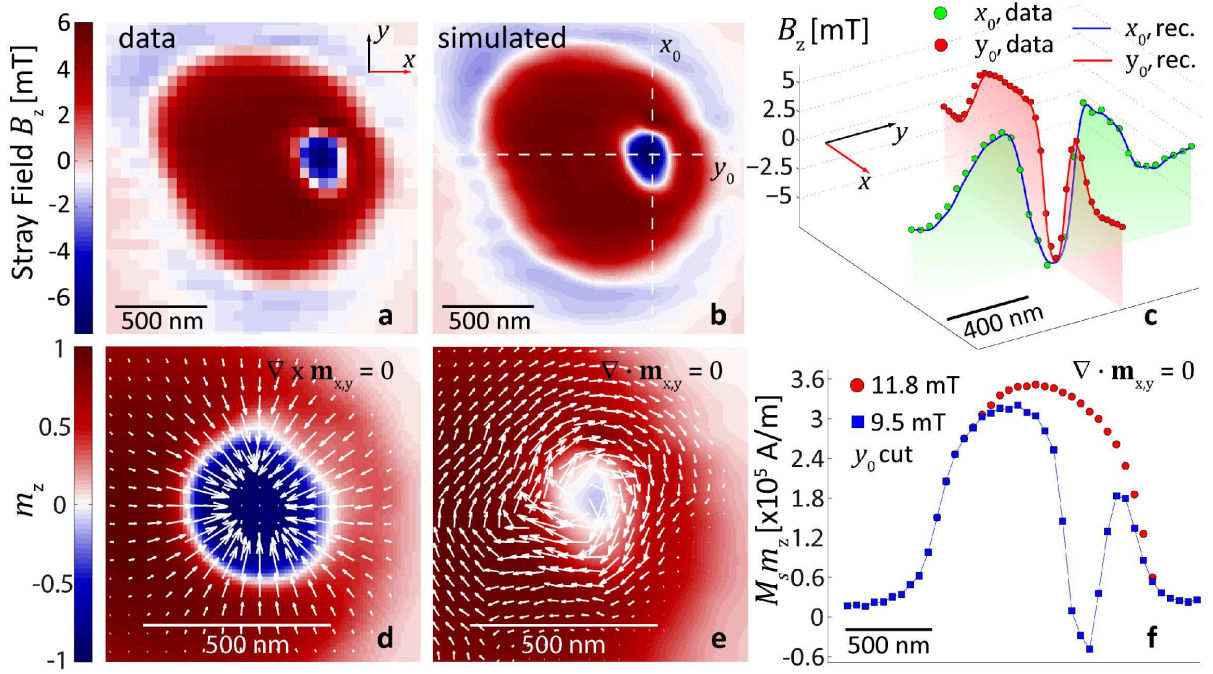}
\vspace{0cm}
\caption{ \textbf{Extracting the local magnetic structure of the skyrmion.} \textbf{a}: $z$-component of the stray field from measured data at a bias field of  $B_{\parallel, \mathrm{ext}}=$9.5 mT applied along the [111] diamond axis. Since a single component of $\mathbf{B}$ contains all relevant information, $B_z$ is chosen for comparison with simulations due to its particularly symmetric coupling to $m_z$ (see text).  \textbf{b}: Simulated map of $B_z$ in both the Bloch and the N\'eel gauge. \textbf{c}: Cuts along the $x=x_0$ and $y=y_0$ lines shown in \textbf{b} (solid lines) and comparison with experimental data in \textbf{a} (markers). \textbf{d}: Magnetic structure obtained in the N\'eel gauge (see Section IIIB in the Supplement). It locally preserves normalization of the local magnetization and produces a stray magnetic field that matches the experimental results. The colour map shows the $m_z$ component. White arrows are proportional to the in-plane magnetization. The deviations of the skyrmion profile from a round shape are most likely related to disc edge effects. \textbf{e}: Plot similar to the one in \textbf{d}, obtained by choosing the Bloch gauge. The local magnetization at the centre of the skyrmion in this case is mostly in-plane. \textbf{f}: Comparison between the reconstructed $M_s m_z$ local magnetization component in the Bloch gauge at two different bias fields (9.5 and 11.8 mT). The $m_z$ profile at saturation (11.8 mT) is used to normalize the local moments for the magnetic structure simulations shown in \textbf{d} and \textbf{e} (see also Section IIIA, IIIB and VII in the Supplement). From this measurement we obtain $M_s m_z \simeq 3.6 \cdot 10^5$ A/m at the disc centre (where $m_z$=1), which agrees with an independently measured value of  $M_s m_z = 3.8 \cdot 10^5$ A/m} 
\label{fig:fig3}
\end{figure} 

A systematic study of the solution manifold requires a way to continuously tune $\gamma$ from the Bloch to the N\'eel case. To vary the helicity, we start by locally rotating the Bloch solution about the $z$-axis by an angle $\lambda(\phi_{\mathrm{N}}-\phi_{\mathrm{B}})$, where $\phi_{\mathrm{N}}$ ($\phi_{\mathrm{B}}$) is the local azimuthal angle of the magnetic structure for the N\'eel (Bloch) configuration. We then perform a rotation about an axis perpendicular to the resulting local moments such as to preserve its in-plane orientation and at the same time match the measured stray field (see Section VI of the Supplement). The parameter $0 \leq \lambda \leq 1$ enables us to move continuously through the manifold. We obtain an ensemble of quantitative, model-independent $m_z(\boldsymbol{\uprho}, \lambda)$ profiles for various values of $\lambda$ as shown in Fig.~4a.\\

In order to select the best candidate texture, we study the topology of the two-dimensional vector field $\mathbf{m}(\boldsymbol{\uprho},\lambda)$.  For any two-dimensional normalized vector field $\mathbf{n}(\boldsymbol{\uprho})$ the topological number $Q$ is defined as:
\begin{equation}
Q = \frac{1}{4 \pi} \int \diff x \diff y \: \mathbf{n} \cdot \left( \frac{\partial \mathbf{n}}{\partial x} \times \frac{\partial \mathbf{n}}{\partial y} \right).
\label{eq:2}
\end{equation}
Whenever $\mathbf{n} \parallel z$ at the boundary, any continuous solution $\mathbf{n}(\boldsymbol{\uprho})$ must have an integer $Q$ value\cite{Ref3}. 
Non-integer values for $Q$ occur in the case of a discontinuity, which is energetically costly and unstable\cite{Ref5}. Meanwhile, skyrmions are stable against local perturbations because of the large energetic cost preventing the skyrmion (Q = ±1) from folding back into the ferromagnetic state (Q = 0). We therefore introduce continuity as a criterion for selecting physically allowed solutions. In Fig.~4b we plot the absolute value of $Q(\lambda)$ for each of the normalized vector fields $\mathbf{n}(\boldsymbol{\uprho},\lambda)$,  with $\mathbf{n}$  being the unit vector in the direction of $\mathbf{m}$. The number $Q$ can be visualised as the number of times the spin configuration $\mathbf{n}$ wraps around the unit sphere\cite{Ref3}. To illustrate the value of Q, in the inset of Fig.~4b we plot the solid angle spanned by $\mathbf{n}$ while moving in the $(x,y)$ plane. We obtain a value for $Q$ approaching -1 as $\lambda \rightarrow 1$. We therefore identify N\'eel or nearly-N\'eel solutions as the only ones compatible with the measured data. \\
\begin{figure}[h!]
\centering
\includegraphics[width=0.72\textwidth]{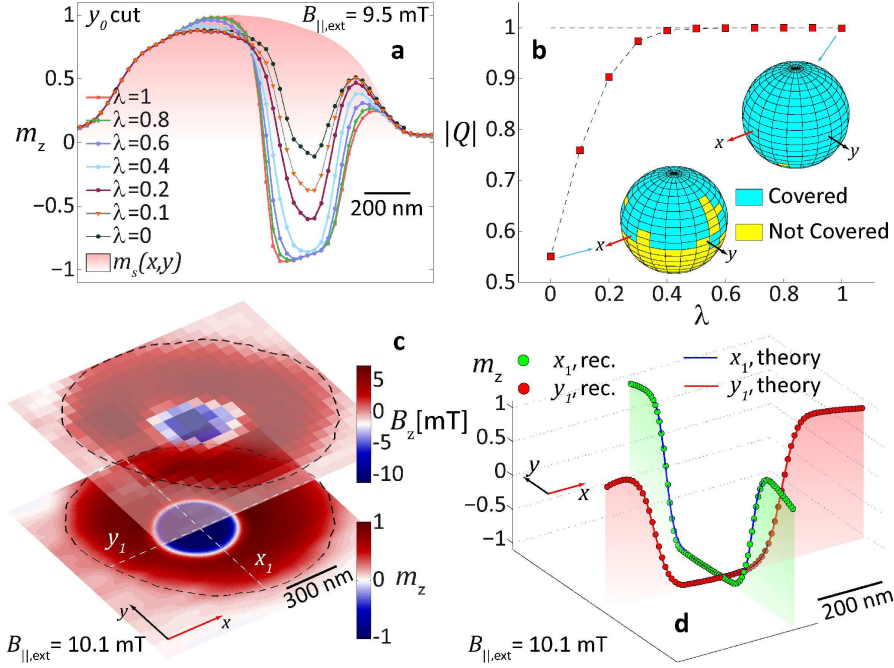}
\vspace{0cm}
\caption{ \textbf{Topology of the reconstructed magnetic structure.} \textbf{a}: Continuous tuning of the magnetic structure from the Bloch to the N\'eel gauge as a function of the parameter $\lambda$ (see text for details). The $m_z$ profiles reported here are cuts along the $y=y_0$ line shown in Fig.~3b. The filled shaded region represents the spatial variation of the normalized saturation magnetization, namely the $m_z$ profile given by the filled red markers in Fig.~3f. \textbf{b}: Absolute value of the topological number defined in eq. \eqref{eq:2}, for each of the spin configurations shown in \textbf{a}. The number $Q$ can be visualized as the number of times the vector field wraps around a unit surface. Therefore, the inset shows the stereographic projection of the vector field on a sphere. The image illustrates that only N\'eel-like configurations have integer $Q$. \textbf{c}: Map of the $B_z$ component of the stray field (upper sheet) and reconstructed $m_z$ magnetization (lower sheet) for a skyrmion nucleated at the centre of the magnetic disc. The black dashed lines represent the disc boundary. The scan was measured with a bias field parallel to the NV axis of $B_{\parallel,\mathrm{ext}}$ =10.1 mT. \textbf{d}: Comparison of the reconstructed $m_z$ skyrmion profile (markers) with a domain wall model for the skyrmion (solid lines). The profiles are cuts through the $x=x_1$ and $y=y_1$ directions shown in Fig.~4c. Spatial variation of the saturation magnetization is taken into account and the skyrmion profile is observed to be round.}
\label{fig:fig4}
\end{figure}
To make a quantitative comparison of our reconstructed $m_z$ profile in the $\lambda=1$ case with analytical expressions, we nucleate another skyrmion in the centre of the disc at a bias field of 10.1~mT along the NV axis (see $B_z$ in Fig.~4c).  The location of this skyrmion minimises possible spurious effects caused by the disc edges and allows us to independently test our reconstruction procedure. When comparing line cuts through the $m_z$ profile at the skyrmion centre with existing models proposed in the literature (see Fig.~4d), we observe an out-of-plane magnetization varying in space as $m_z(\tilde{\rho}) = \tanh \left( \frac{\tilde{\rho} - \rho_0}{w/2} \right)$, with $\rho_0$ and $w$ being the skyrmion radius and domain wall width and with $\tilde{\rho}$ being the distance from the skyrmion centre\cite{Ref29}.  Our helicity and $m_z$ shape are in agreement with the recent first high spatial resolution skyrmion images by X-ray magnetic circular dichroism microscopy and spin-resolved STM at low temperature\cite{Ref11,Ref29}. The NV-to-film distance $d\sim$30 nm is too large to extract the domain wall width $w$, but it is sufficient to determine the skyrmion radius $\rho_0 \simeq$ 210 nm for the cross sections along the $(x_1,y_1)$ directions shown in Fig.~4d at $B_{\parallel,\mathrm{ext}}=$10.1 mT.\\

Our analysis consistently identifies right-handed ($\gamma = \pi$) N\'eel-like skyrmions as the only continuous solutions with fixed helicity if we require that the structure does not vary through the sample thickness. N\'eel skyrmions are expected from theory when surface inversion symmetry leads to a Rashba-type DMI \cite{Ref17} and the latter dominates over magnetostatic contributions\cite{Ref9}. However, the expected chirality is left-handed ($\gamma = 0$), based on recent X-ray magnetic circular dichroism microscopy measurements of single Pt/Co layers in zero field \cite{Ref11}, indirect transport measurements in Pt/Co multilayers through skyrmion movement\cite{Ref9}, and studies of domain walls in Pt/Co\cite{Belmeguenai, Emori, Pizzini, Ryu}, reporting $\gamma = 0$. In contrast with previous data, our skyrmions are not left-handed.\\

Helicity is dictated  by the nature of the energy terms resulting from the breaking of the spatial inversion symmetry along the $z$-axis. In the absence of DMI, 
Bloch ($\gamma= \pm \pi/2$) configurations are expected\cite{Montoya}. The presence of a chiral DMI term produces $\gamma=0$ configurations \cite{Ref11}. For thick multilayer dots, ev
en with no DMI the magnetic layers in the vicinity of the top (bottom) surface will experience a breaking of the $z\rightarrow-z$ inversion symmetry, favouring N\'eel spin textures with right-handed (left-handed) chirality \cite{Montoya}. Such twisted structures (also known as N\'eel caps) reduce the stray field and accordingly the demagnetization energy cost. N\'eel caps would not be visible with techniques averaging over the sample thickness, such as Lorentz TEM \cite{Montoya, Ref19}. Our technique is most sensitive to the topmost layer, thus our observation of a right-handed skyrmion is the first to indicate the presence of a N\'eel cap.\\
\begin{figure}[h!]
\centering
\includegraphics[width=1\textwidth]{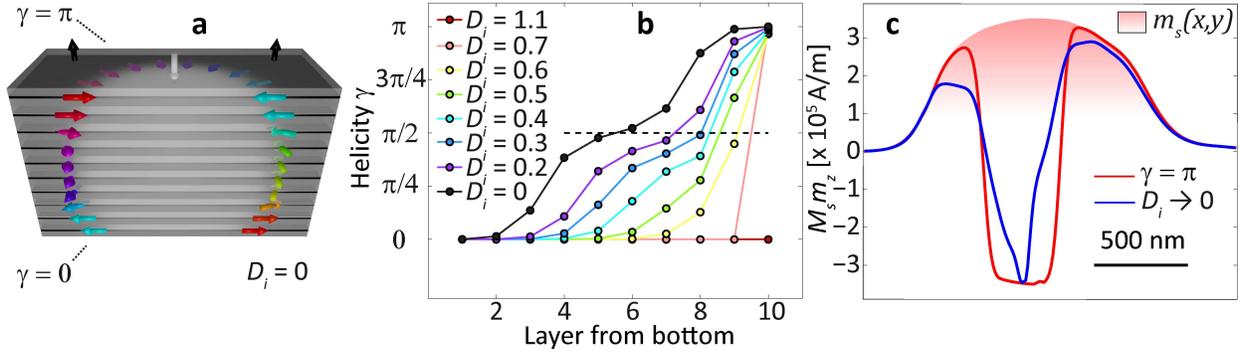}
\vspace{-0.5cm}
\caption{ \textbf{N\'eel caps in magnetic multilayers hosting topological spin structures. a}: Sketch of the magnetic texture obtained via a micromagnetic numerical simulation. The closure domains (i.e. N\'eel caps \cite{Montoya}) at the top and bottom of the multilayer reduce the demagnetization energy cost with respect to the purely Bloch case. In the simulation $M_s= 10^6~\mathrm{A/m}$, $A= 10~\mathrm{pJ/m}$, magnetic anisotropy field is 0.2~T, and $D_i=0$ (see Section IX of the Supplement). The number of layers and separation is representative of the measured sample. The non-uniformity of $M_s$ and layer thicknesses is not taken into account for this simulation, which may lead to an underestimation of dipolar effects. \textbf{b}: Local helicity for each one of the 10 magnetic layers as the DMI value is varied. The DMI is expressed in $\mathrm{mJ/m^2}$. Skyrmions with $\gamma\rightarrow\pi (\gamma\rightarrow0)$ are present at the top (bottom) of the stack.  \textbf{c}: Cut through the reconstructed $m_z$ profiles from topologically protected textures that produce a stray field matching the experimental data in Fig. 4c. The red curve corresponds the effective gauge fixed at $\gamma=\pi$ for each layer; the blue curve corresponds to a value of $\gamma=\pi(\gamma=0)$ for the top (bottom) three layers, and $\gamma = \pi/2$ for the four layers in the middle. This red curve approximates the $D_i\rightarrow0$ case depicted in panel \textbf{a}. The filled shaded region represents the spatial variation of the saturation magnetization. The NV depth was again fixed at 30 nm.}
\label{fig:fig5}
\end{figure}
In order to test the energetic stability of skyrmions with changing helicity through the sample thickness, we ran micromagnetic simulations of ten representative proximal magnetic layers, for simplicity with spatially uniform microscopic energy terms (see Fig. 5 and details in section IX of the Supplement). In the limiting case of no DMI ($D_i\rightarrow0$), the top and bottom layers have opposite N\'eel chiralities, while the intermediate layers are Bloch-like (see Fig. 5a). For small values of the DMI term $D_i$ (see Fig. 5b), right-handed skyrmions are stabilized within the top layers. In order to attempt a comparison of the structure in Fig. 5a with the measured data we look for a solution with an effective gauge varying through the sample thickness, which is N\'eel-like for the top and bottom three layers and Bloch or Coulomb-like for the central part of the multilayer (see section X of the Supplement for the details of this procedure). By numerically minimizing the difference between measured and computed field (See Sections III and IV of the Supplement), we obtain the local $m_z$ profile represented by the blue line in Fig. 5c. We compare this solution with the skyrmion solution previously obtained in Fig. 4d (solid red line). The new $z-$dependent solution still satisfies $Q\rightarrow -1$, but its $m_z$ profile is less sharp. We believe that this shape is due to the variation in skyrmion radius across the multilayer thickness, as suggested by simulations (see e.g. Fig. 5a). 
The presence of N\'eel caps and small DMI thus reconciles our data with recent reports of left-handed structures in multilayers and provides evidence in favor of a previously unobserved phenomenon in these films.\\

In the broader perspective, our work is the first  example of full vector magnetometry and spin reconstruction performed with a single NV centre. It also provides an answer to the long-standing magnetometry problem of reconstructing the full set of spin textures from a measured stray field, using a general formalism readily applicable to all local magnetometry techniques.  The crucial advantage of our technique is its locality and enhanced sensitivity to the topmost magnetic layers. Here, we applied these methods to N\'eel caps in magnetic skyrmions hosted in sputtered Pt(3~nm)/Co(1.1~nm)/Ta(4~nm) stacks. In contrast with previous work, we rule out purely left-handed N\'eel solutions in magnetic multilayers. We show that our results are consistent with a previously unobserved twisted structure with vertically evolving chirality and helicity, which is expected from micromagnetic simulations. Our results and methods will be broadly relevant to nanoscale magnetometry and studies of chiral spin textures for room-temperature spintronics applications \cite{Ref5,Ref9,Ref14}, as well as imaging of current distributions\cite{Ref20, Chang} and magnetic structures in low-dimensional materials\cite{Huang}.\\


\textbf{Methods}\\
\textit{Sample fabrication and measurement protocol:} Magnetic discs are patterned on the flat surface of a cleaved quartz tip, pictured in Fig.~1c, by electron beam lithography (see Section I of the Supplement). The quartz tip is then mounted on a piezo-electric tuning fork. Monitoring the resonance frequency of the fork allows us to maintain a constant force between the sample and the pillar\cite{Ref15}. We choose the quartz tip diameter to be $\sim$50 $\upmu$m, which allows us to selectively approach an individual NV pillar chosen from a grid of pillars spaced by 50 $\upmu$m and fabricated on a 2x4 mm diamond wafer. We deposit a coplanar waveguide (CPW) on the surface of the diamond, aligned in such a way that rows of pillars reside in gaps. The CPW is used for driving NV centre spin transitions. Optical addressing of the NV centre is done through the 50 $\upmu$m thick diamond.  The green laser power used for optical excitation of the NV centre is $\sim$100 $\upmu$W, reduced well below optical saturation in order to avoid heating the sample.  A bias magnetic field is delivered by a permanent magnet mounted on a mechanical stage. The magnetic field is aligned parallel to the NV axis, following a procedure based on the NV photoluminescence\cite{Ref26}. This allows us to measure the evolution of magnetic features as function of applied external field, with the field pointing along the NV axis. The nominal value of $M_s$ for the Pt/Co/Ta multilayer film is independently measured using a reference sample placed in the sputtering chamber together with the quartz tip during the deposition process and is found to be $M_s m_z = 3.8 \cdot 10^5$ A/m (see Section VII of the Supplement).\\

\textbf{Acknowledgements}\\
This work is supported by the Gordon and Betty Moore Foundation’s EPiQS Initiative through Grant GBMF4531. A.Y. and R.L.W. are also partly supported by the QuASAR and the MURI QuISM projects. Work at MIT was supported by the U.S. Department of Energy (DOE), Office of Science, Basic Energy Sciences (BES) under Award no. DE-SC0012371 (sample fabrication and magnetic properties characterization). F.C. acknowledges support from the Swiss National Science Foundation (SNSF) grant no. P300P2-158417. S.S. acknowledges the National Science Foundation Graduate Research Fellowship under grant no. DGE1144152. F.B. acknowledges financial support by the German Research Foundation through grant no.  BU 3297/1-1. Diamond samples were provided by Element Six (UK). We thank Dr. Marc Warner (Harvard) for helpful ideas in the initial stages of the experiment and Dr. Rainer St\"ohr (Harvard - Stuttgart) for technical advice.  We thank James Rowland (Ohio State) for fruitful discussions.\\

\textbf{Author Contributions}\\
Y.D., F.C., S.S., G.S.D.B., and A.Y. conceived the experiment. T.Z. and F.C. designed and developed the quartz tips and the diamond. T.Z. optimized the fabrication procedure. S.S. developed the deposition recipes and optimized the magnetic properties of the multilayers. Y.D. and F.C. performed the experiment. F.C. developed the theoretical model. F.C. and Y.D. performed data analysis. R.L.W., G.S.D.B. and A.Y. provided guidance. F.B. and G.S.D.B proposed the twisted skyrmion model, and F.B. carried out the associated simulations. A.Y. supervised the work. All authors contributed to the writing and the content of the manuscript.
\\
\begingroup
\renewcommand{\addcontentsline}[3]{}

\endgroup

\clearpage
\widetext
\begin{center}
\textbf{\Large Supplementary Information:\\ Imaging the Spin Texture of a Skyrmion Under Ambient Conditions Using an Atomic-Sized Sensor}
\end{center}
\setcounter{equation}{0}
\setcounter{figure}{0}
\setcounter{table}{0}
\setcounter{page}{1}
\makeatletter

\tableofcontents

\section{Sample preparation}
\subsection{Geometry of the scanning probe experiment}
\noindent

In our experiment, magnetic discs were deposited at the top of cleaved quartz fiber~$\sim$~50~$\upmu$m in diameter. 
We employ Sutter Instruments quartz rods with the initial diameter of 1~mm. The fiber diameter is controlled via a laser-based puller (P-2000  Sutter Instruments). Fabrication proceeds with a controlled mechanical cleaving of the pulled fiber after the introduction of an intentional crack by means of a diamond scribe.
\begin{figure}[h!]
\centering
\includegraphics[width=0.8\textwidth]{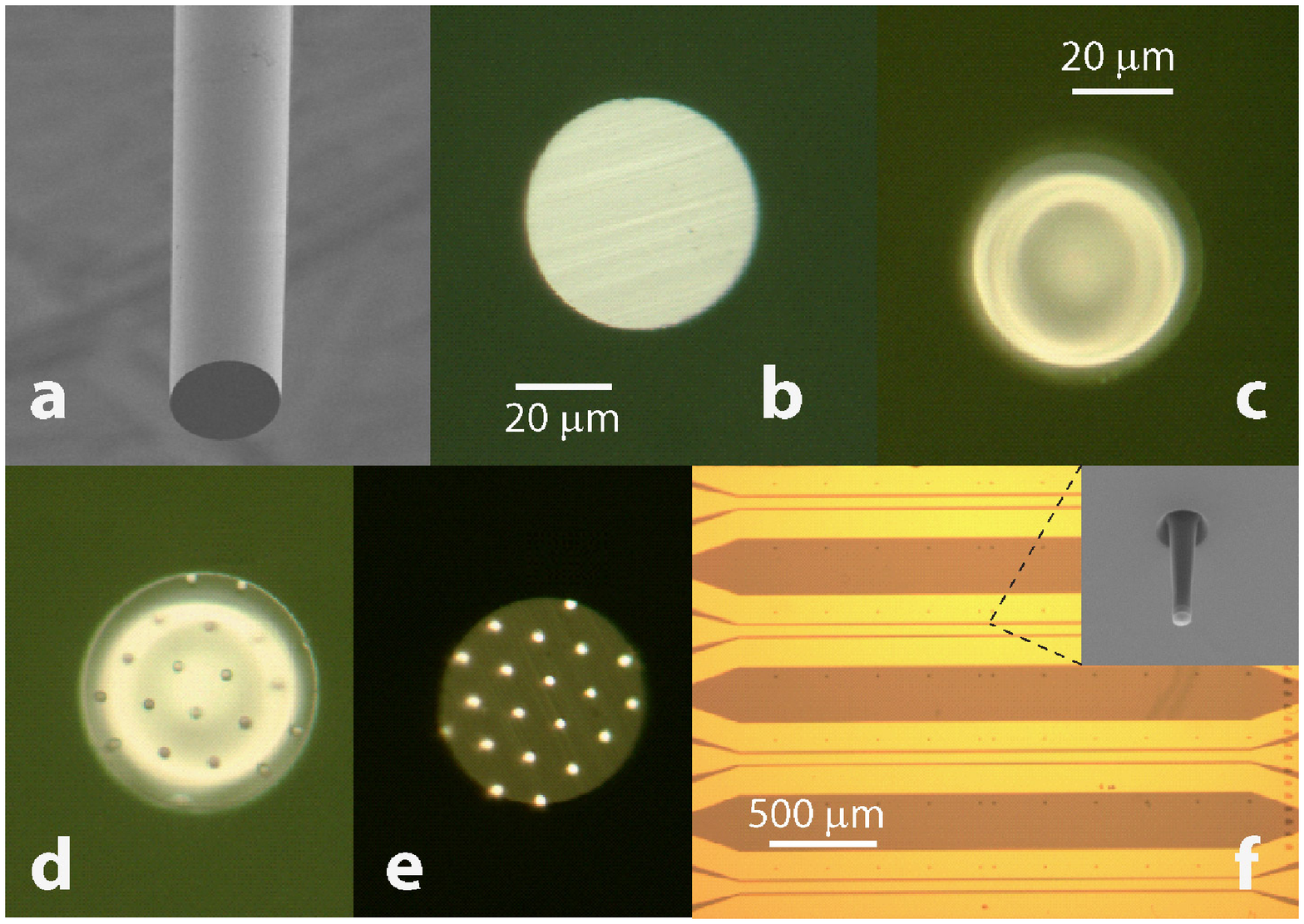}
\caption{\textbf{Fabrication of the tip with magnetic discs.} \textbf{a:} Electron microscopy image of a pulled quartz fiber after the mechanical cleaving. \textbf{b:} Optical microscopy image of the top surface of the tip, after being glued face-up on the aluminium holder. \textbf{c:} The quartz tip after being spin coated with the PMMA resist. \textbf{d:} A triangular lattice array of discs having a 2~$\upmu$m  diameter is defined using electron-beam lithography. \textbf{e:} The tip after lift-off. The sputtered Pt/Co stacks are visible as highly reflective dots on the tip surface.  \textbf{f:} Four adjacent Ti:Au (5:100~nm) coplanar waveguides deposited via photolithography on the diamond surface. Inset: rows of diamond pillars are present within the gap between the waveguide plates.}
\label{fig:fab}
\end{figure}
The resulting tip, shown in Fig.~\ref{fig:fab}a, is glued face-up (Fig.~\ref{fig:fab}b) on an aluminum holder (not shown). The position of the tip with respect to markers present on the holder can be measured via electron microscopy prior to the application of the resist used for lithography. 
The holder is subsequently mechanically mounted on a conventional spinner disc. We applied on the free-standing tips a few drops of a Microchem C6 PMMA resist. After spinning, the deposited resist (see Fig.~\ref{fig:fab}c) is exposed via electron-beam lithography with a triangular lattice array of discs having a 2~$\upmu$m  diameter (Fig.~\ref{fig:fab}d).  We then deposited via sputtering a Ta(3~nm)/[Pt(3~nm)/Co(1.1~nm)/Ta(4~nm)]x10 stack. 
The bottom Ta layer is deposited as a seed layer to increase sample adhesion and enhance perpendicular magnetic anisotropy. The metal layers are deposited by d.c. magnetron sputtering at 2 mTorr Ar (Ta) and 3 mTorr Ar (Pt, Co), with a background pressure of $5 \cdot 10^{-6}$ Torr. Deposition rates are $<0.1$~nm~s$^{-1}$ and calibrated by X-ray reflectivity. Reference Si substrates are held at the same height as the surface of the quartz tips and used to calibrate the saturation magnetization of the material (see also Section \ref{sec:calib}).
Subsequent immersion into acetone of the tip and the holder dissolves the glue and the resist, leaving the tip in the final state shown in Fig.~\ref{fig:fab}e, ready to be glued onto the tuning fork used for the experiment.
%
\subsection{Diamond fabrication}
\noindent
Our experiments were performed with a type IIa diamond grown by chemical vapor deposition by Element 6 measuring 4x2x0.05~mm$^3$. We studied NV centres formed by N$^{15}$ ion implantation at an energy of 18 keV and a density of 500/$\upmu$m$^2$ and subsequent annealing for 2 hours at 800$^\circ$C. This implantation energy is expected to yield NV centres at an estimated $\sim$30~nm depth from the diamond surface \cite{Spinicelli11}.\\
For the diamond pillars we first prepared an etch mask, patterned on the diamond via electron beam lithography using a FOX 16 flowable oxide resist from Dow Corning. Adhesion of the resist was guaranteed by a very thin, $\sim$ 10 nm layer of Titanium deposited via electron beam evaporation.  The exposed etch mask pattern was transferred onto the diamond via a conventional top down anisotropic plasma etch performed in a Unaxis Shuttleline ICP reactive ion etching (RIE) system.
An initial Ar/Cl$_2$ plasma etch was used to remove the Titanium adhesion layer, while O$_2$ plasma was used to etch the diamond. A typical $\sim$ 1.5~$\upmu$m tall and $\sim$ 200~nm wide diamond pillar, imaged via electron microscopy, is shown in the inset of Fig.~\ref{fig:fab}f. We then defined a set of Ti:Au (5:100~nm) coplanar waveguides via photolithography. The gaps between the central conductor and the ground plate of the waveguides were aligned (see Fig.~\ref{fig:fab}f) with the pillar rows via alignment markers defined on the diamond during the same O$_2$ etching described above.
\section{Principles of stray field magnetometry}
\subsection{Sensing static fields with an NV centre}
Because of the spatial confinement of its local spin density, to a volume below 1~nm$^3$ (see Ref. \onlinecite{Gali08}), a nitrogen-vacancy centre in diamond can be well approximated as a point-like sensor of magnetic fields. \\
Following other works \cite{Balasubramanian08,Casola15}, we will consider the following Hamiltonian for this spin-1 defect:
\begin{equation}
\mathscr{H} = D(\hat{S_{\parallel}})^2 + \gamma B_{\parallel} \hat{S_{\parallel}} + \gamma B_{\perp} \hat{S_{\perp}},
\label{eq:HaminField}
\end{equation}
where $D$ is the zero-field splitting, $\gamma = 2.8025$ MHz/G is the NV gyromagnetic ratio and $\parallel, \perp$ indicate the directions parallel and perpendicular to the spin quantization axis of the colour centre. In this work, our [100]-cut diamond hosts NV centres forming an angle $\theta_{\mathrm{NV}} = \arccos(1/\sqrt{3})$ with the surface normal, defining the $\parallel$ direction (see also Fig.~\ref{fig:CoordRes}). \\
The value for $D$ was obtained from a measurement of the electron-spin resonance (ESR) line splitting at small applied fields and was found to be $D = 2.8710(1)$ GHz.\\
In our experiments, both the upper ($\omega_{+}$) and lower ($\omega_{-}$) NV resonance frequency was measured at each point in space. The values for $B_{\parallel}$ and $B_{\perp}$ were then obtained using the following expressions \cite{Casola15}:
\begin{align}
&B_{\parallel} = \frac{\sqrt{-(D + \omega_+ - 2\omega_-)(D+\omega_--2\omega_+)(D+\omega_-+\omega_+)}}{3\gamma\sqrt{3D}}, \label{eq:hz} \\
&B_{\perp} = \frac{\sqrt{-(2D-\omega_+-\omega_-)(2D+2\omega_--\omega_+)(2D-\omega_-+2\omega_+)}}{3\gamma\sqrt{3D}}. \label{eq:hx}
\end{align}
\subsection{Reconstruction of the stray field components}
\label{sec:rec}
In our experiment we measure the stray field $B_{\parallel}(x,y)$ in a plane at a distance $d$ from the magnetic surface. We can call such quantity $B_{\parallel}(\bm{\uprho},d)$, with $\bm{\uprho} = (x,y)$. \\
At the probe position the stray field is curl free and it is therefore possible to define a magnetostatic potential $\phi_{\mathrm{M}}$ such that the vector field can be written as \cite{Blakely96,Dreyer07}:
\begin{equation}
\mathbf{B} = - \nabla \phi_{\mathrm{M}}(x,y).
\end{equation}
As pointed out in the past\cite{Lima09,Blakely96}, the previous relation implies that the stray field components are not independent. We will now derive the relation among the different stray field components and obtain further insight by starting from the following 2D Fourier transform of the quantity $\mathbf{B}(\bm{\uprho},d)$, defined in Ref. \onlinecite{Casola15} as:
\begin{align}
\mathbf{B}(\mathbf{k},d) = \int_{0}^{2 \pi} \int_{0}^{\infty} \mathbf{B}(\bm{\uprho},d) e^{-i k {\rho} \cos(\phi-\phi_k)}  {\rho} \diff {\rho} \diff \phi. 
\label{eq:kernel}
\end{align}
\begin{figure}[h!]
\centering
\includegraphics[width=\textwidth]{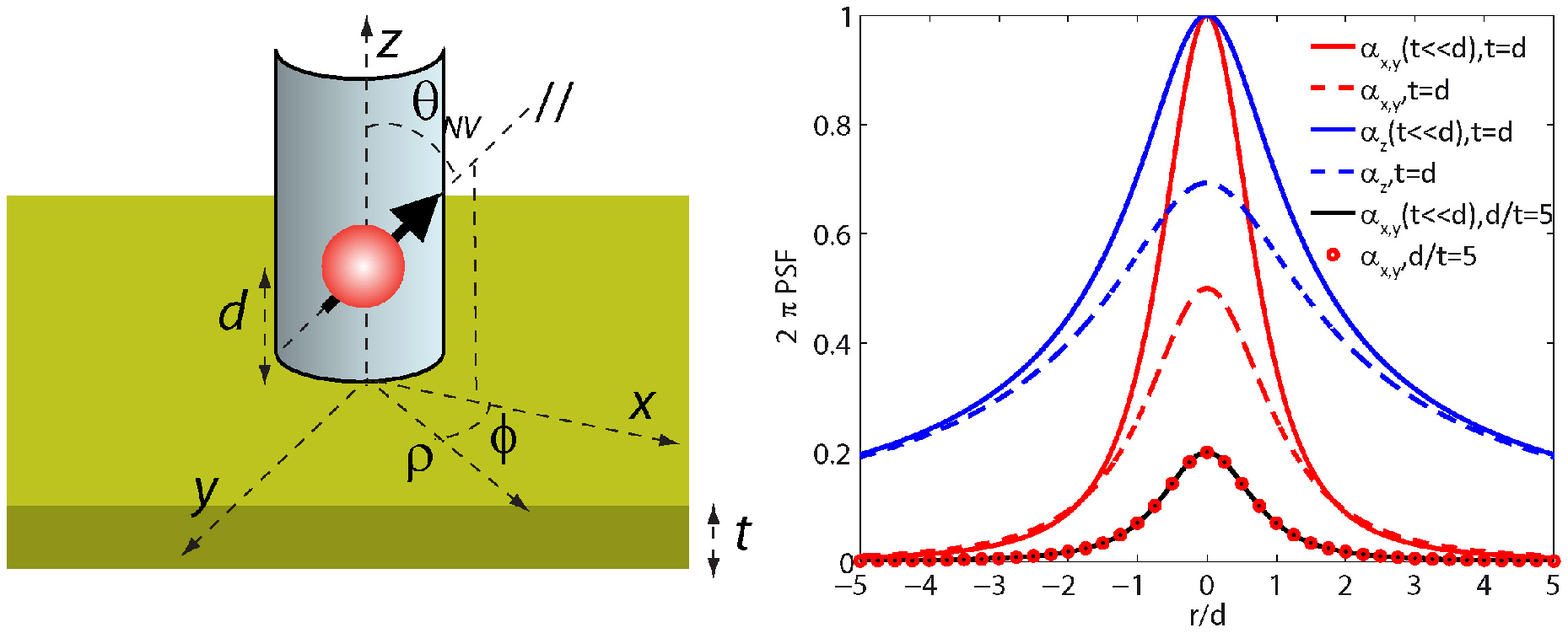}
\caption{\label{fig:CoordRes} \textbf{Universal resolution functions for NV magnetometry} \textbf{Left:} Reference frame used for the calculations discussed in the text. The sketch represents an NV in a [100]-cut diamond pillar, with a quantization axis at an angle of $\theta_{\mathrm{NV}} = 54.7^{\circ}$ with respect to the surface normal. The in-plane projection of the NV quantization axis is parallel to the $x$-axis. The NV has a distance $d$ from the surface of the magnetic material and the magnetic film has a thickness $t$. \textbf{Right:} In the figure we plot the resolution functions derived in Section \ref{sec:alphas} for various $t/d$ ratios. Note that the resolution functions $\alpha_{x,y}$ and $\alpha_z$ have different units. In this plot the space dependence is shown with all the point spread functions (PSFs) plotted using $d=1$. For the regime $t \ll d$ we use equations \eqref{eq:res} and \eqref{eq:res2}, while the full expression for $\alpha_{x,y}, \alpha_z$ is retained in all the other cases.}
\end{figure}
The vectors $\textbf{k}$ and $\bm{\uprho}$ are 2-dimensional vectors in reciprocal and real space, forming an angle $\phi_k$ and $\phi$ with the $x$-axis (see also left of Fig.~\ref{fig:CoordRes}). As derived in Ref.\ \onlinecite{Casola15}, when the stray field is produced by a sheet of magnetic dipoles distributed over a thickness $t$ and with local magnetization $M_s \mathbf{m}(\bm{\uprho}_j) = M_s [m_x(\bm{\uprho}_j),m_y(\bm{\uprho}_j),m_z(\bm{\uprho}_j)]$, the stray field can be written in momentum space as \cite{Casola15}:  
\begin{equation}
\mathbf{B}(\mathbf{k},d) = \mathbf{D}(\mathbf{k},d) \mathbf{m}(\mathbf{k}),
\label{eq:FeidlLab}
\end{equation}
where the expression for the traceless symmetric $\mathscr{D}(\mathbf{k},d)$ kernel matrix reads:
\begin{align}
\mathbf{D}(\mathbf{k},d) = \frac{\upmu_0 M_s}{2} (e^{-dk} - e^{-(d+t) k}) \left( \begin{array}{ccc}
 - \cos^2(\phi_k) & - \frac{\sin(2\phi_k)}{2} & -i \cos(\phi_k) \\
 - \frac{\sin(2\phi_k)}{2} & - \sin^2(\phi_k) & -i \sin(\phi_k) \\
 -i \cos(\phi_k) & -i \sin(\phi_k) & 1 \end{array} \right).
\label{eq:dipoDefKernel}
\end{align}
Note that with respect to Ref. \onlinecite{Casola15}, we have included the finite size film thickness $t$ by assuming the local magnetization vector $\mathbf{m}(\bm{\uprho}_j)$ to be constant through the magnetic film thickness and therefore integrating that dimension out. $M_s$ is the nominal, space-independent, saturation magnetization of the magnetic film. \\

\noindent
From eq. \eqref{eq:dipoDefKernel} we can immediately realize that the rows of the matrix $\mathbf{D}(\mathbf{k},d)$ are not independent and for this reason $\mathbf{D}(\mathbf{k},d)$ is not invertible. In general, in eq. \eqref{eq:FeidlLab} it is impossible to obtain $\mathbf{m}$ by simply measuring all the components of the vector $\mathbf{B}$.\\
In momentum space, the algorithm relating the stray field component along the $z$-axis $B_z(\mathbf{k},d)$ (see Fig.~\ref{fig:CoordRes}) to $B_{\parallel}(\mathbf{k},d)$, for an NV lying in the $zx$-plane, can be simply written as:
\begin{equation}
B_z(\mathbf{k},d) = \frac{B_{\parallel}(\mathbf{k},d)}{\cos(\theta_{NV}) - i \sin(\theta_{NV}) \cos(\phi_k)},
\label{eq:reco1}
\end{equation}
In a similar way, all vector field components can be reconstructed without singularities from a single measurement of $B_{\parallel}(\bm{\uprho},d)$ within a whole plane, provided $\theta_{NV} \neq \pi/2$.\\ 
Finally, we point out that the expression in eq. \eqref{eq:dipoDefKernel} includes also an analogy with the Huygens principle in optics. In particular, in order to reconstruct the field within a plane at a different distance $d' = d + h$ from the film, it will be simply enough to perform an inverse Fourier transform of the 2D Fourier transform at a distance $d$, multiplying by the prefactor $\exp(-k h)$. Such operations are known as upward or downward propagation for $h>0$ or $h<0$ and are well discussed in the literature \cite{Blakely96}.\\
Once $B_z$ is known, all the other field components can be reconstructed according to \eqref{eq:dipoDefKernel}. For instance, $B_x(\mathbf{k},d) = - i \cos(\phi_k) B_z(\mathbf{k},d)$. Using full knowledge of all the stray vector field components, in Fig.~2 of the main text we have therefore reconstructed the expected magnitude of the stray field transverse to the NV axis and originating from the magnetic disc, using the expression:
\begin{equation}
B_{\perp,r}(\bm{\uprho},d) = \sqrt{B_y^2 + (B_z \sin(\theta_{\mathrm{NV}}) - B_x \cos(\theta_{\mathrm{NV}}))^2}.
\label{eq:transverseField}
\end{equation}
Note that the reconstructed map of $B_{\perp,r}(\bm{\uprho},d)$ will exactly match the map extracted from the spin level mixing given knowledge of the uniform bias field, which in our case is known up to the direction of a small perpendicular component.

\subsection{Real space interpretation and effective point spread function of NV magnetometry}
\label{sec:alphas}
As we shall see, eq. \eqref{eq:dipoDefKernel} allows for an intuitive real-space interpretation.\\ 
We define the real-space expression for the \textit{resolution function} $\alpha_{x,y}(d,t)$ of the in-plane  magnetization as:
\begin{align}
& \alpha_{x,y}(d,t) = \frac{1}{(2 \pi)^2} \int_{\mathbf{k}}  \frac{e^{-(d+t) k}\left( e^{t k} - 1 \right)}{k} e^{i \mathbf{k} \cdot \bm{\uprho}} \diff \mathbf{k} \nonumber \\
& = \frac{1}{2 \pi} \left(  \frac{1}{\sqrt{d^2 + r^2}} - \frac{1}{\sqrt{(d+t)^2 + r^2}}	\right).
\end{align}
Note that if $t \ll d$ the previous resolution function can be simplified as:
\begin{align}
\alpha_{x,y}(d,t \ll d)  \approx \frac{1}{2 \pi} \frac{d t}{(d^2 + r^2)^{3/2}}.
\label{eq:res}
\end{align}
In the same way, we define the real-space expression for the \textit{resolution function} $\alpha_{z}(d,t)$ of the out-of-plane magnetization as:
\begin{align}
& \alpha_{z}(d,t) = \frac{1}{(2 \pi)^2} \int_{\mathbf{k}}  \frac{e^{-(d+t) k}\left( e^{t k} - 1 \right)}{k^2} e^{i \mathbf{k} \cdot \bm{\uprho}} \diff \mathbf{k} \nonumber \\
& = - \frac{1}{(2 \pi)^2} \int_{d'} \int_{\mathbf{k}}  \frac{e^{-(d'+t) k}\left( e^{t k} - 1 \right)}{k} e^{i \mathbf{k} \cdot \bm{\uprho}} \diff \mathbf{k} \diff d' \nonumber \\
& = - \int \alpha_{x,y}(d',t) \diff d' =  \frac{1}{2 \pi} \log \Big( \frac{ d + t + \sqrt{(d+t)^2 + r^2}}{d + \sqrt{d^2 + r^2}} \Big).
\end{align}
Once more, if $t \ll d$ the previous resolution function can be simplified as:
\begin{align}
\alpha_{z}(d,t \ll d)  \approx \frac{1}{2 \pi} \frac{t}{(d^2 + r^2)^{1/2}}.
\label{eq:res2}
\end{align}
The spatial dependence of these resolution functions or effective \textit{point spread functions} (PSFs) for magnetometry is plotted to the right of Fig.~\ref{fig:CoordRes} for different sets of parameters.\\ 
With such notation, no approximations, and using the convolution theorem we can rewrite the real space expression for $\mathbf{B}(\bm{\uprho},d)$ as:
\begin{align}
& \mathbf{B}(\bm{\uprho},d) =  - \frac{\upmu_0 M_s}{2} \left( \begin{array}{ccc}
- \alpha_{z}(d,t) \ast \frac{\partial^2  }{\partial x^2}		& - \alpha_{z}(d,t) \ast \frac{\partial^2  }{\partial y \partial x}		& 	\alpha_{x,y}(d,t) \ast \frac{\partial}{\partial x}	\\
- \alpha_{z}(d,t) \ast \frac{\partial^2  }{\partial y \partial x}		&	- \alpha_{z}(d,t) \ast \frac{\partial^2  }{\partial y^2}		& 	\alpha_{x,y}(d,t) \ast \frac{\partial}{\partial y}	\\
\alpha_{x,y}(d,t) \ast \frac{\partial}{\partial x}		&		\alpha_{x,y}(d,t) \ast \frac{\partial}{\partial y}	&		\alpha_{z}(d,t) \ast \nabla^2	 \end{array} \right) \left( \begin{array}{c} m_x(\bm{\uprho}) \\ m_y(\bm{\uprho}) \\ m_z(\bm{\uprho}) \end{array} \right).
\label{eq:dipoDefKernelApp2}
\end{align}

\noindent
We have discussed in Section \ref{sec:rec} the fact that a single component of the stray magnetic field vector carries all the information, as all the other components are fixed given the first one. Due to the symmetry of our problem it is particularly illuminating to consider the B$_z$ component:
\begin{equation}
B_z(\bm{\uprho},d)= - \frac{\upmu_0 M_s}{2} \left( \alpha_{z}(d,t) \ast \nabla^2 m_z(\bm{\uprho})  + \alpha_{x,y}(d,t) \ast \nabla \cdot \mathbf{m}_{x,y}(\bm{\uprho}) \right),
\label{eq:Bz}
\end{equation}
with $\mathbf{m}_{x,y} = (m_x,m_y)$ the in-plane magnetization vector. 
It's clear that \eqref{eq:Bz} contains convolutions and it therefore entails the non-locality of the dipolar tensor. 
At the same time, the two resolution functions $\alpha_{x,y}$ and $\alpha_{z}$ for the in-plane and out-of-plane component of the magnetization are not equal. An intuitive reason is given by the analogy between magnetic moments and current distributions. When the magnetization is out of plane, the stray field can be viewed as given by an effective current flowing at the boundaries of the region of constant $m_z$. This means that the magnetic field scales as $\sim 1/r$, being $r$ the distance from the source. For the in-plane magnetization case the situation is instead equivalent to two current sheets above and below the magnetic film; in far field such sheets compensate each other much faster than $\sim 1/r$ and more like an isolated dipole of the form $\sim 1/r^3$. It is evident from \eqref{eq:res} and \eqref{eq:res2} that $\alpha_{z}$ and $\alpha_{x,y}$ indeed scale as $\sim 1/r$ and as $\sim 1/r^3$ for very thin films.\\
Finally, in our experiments we have considered a stack of $N=10$ magnetic thin films separated by a distance $s$. As the magnetization $M_s \mathbf{m}$ is assumed to be constant through the film thickness, the stray field for the $N \neq 1$ case would read exactly like \eqref{eq:Bz} with the difference that the PSFs are replaced by:
\begin{align}
& \alpha_{z}(d,t) \rightarrow \alpha_{z,N}(d,t) = \sum_{\nu =0}^{N-1} \alpha_{z}(d + \nu\cdot s,t), \nonumber \\ 
& \alpha_{x,y}(d,t) \rightarrow \alpha_{x,y,N}(d,t) = \sum_{\nu =0}^{N-1} \alpha_{x,y}(d + \nu\cdot s,t). 
\label{eq:res5}
\end{align}
\section{Magnetization reconstruction in the \textit{Bloch} and \textit{N\'eel} gauge}
\label{sec:gaugeFix}
In the previous Section, eq. \eqref{eq:Bz} provided us with a real space interpretation of stray field magnetometry. As convolutions commute with derivatives, we can reformulate the problem of reconstructing the underlying magnetization pattern from the stray field measurements starting from Gauss's equation:
\begin{equation}
B_z(\bm{\uprho},d)= - \nabla \cdot \mathbf{F},
\label{eq:gauss}
\end{equation}
where the two-component vector field $\mathbf{F}(\bm{\uprho},d)$ plays the role of an effective electric field and the function $B_z(\bm{\uprho},d)$ describes the charge density.\\ 
The effective electric field can be written down as:
\begin{equation}
\mathbf{F}=  \frac{\upmu_0 M_s}{2} \left( \alpha_{z}(d,t) \ast \nabla m_z(\bm{\uprho})  + \alpha_{x,y}(d,t) \ast  \mathbf{m}_{x,y}(\bm{\uprho}) \right),
\label{eq:F}
\end{equation}
where $m_z(\bm{\uprho})$ and $\mathbf{m}_{x,y}(\bm{\uprho})$ play the role of an effective scalar and vector potential, respectively.
A solution to \eqref{eq:gauss} is defined up to a divergenceless term, which in our case can be written as:
\begin{equation}
\mathbf{F} = - \nabla V + \nabla \times C_z  {\mathbf{u}_z},
\label{eq:effeF}
\end{equation}
where $C_z(\bm{\uprho},d)$ is an arbitrary function of space, a priori undetermined, and ${\mathbf{u}_z}$ a unit vector perpendicular to the surface. We choose $C_z(\bm{\uprho},d)\mathbf{u}_z$ to point in the $z$  direction because \textbf{F} is oriented in the $(x,y)$ plane.
Our derivation diverges from classical electromagnetism (EM) \cite{Griffith}. In particular, in EM the curl of the vector potential is determined by a magnetic field measurement. In our effective problem we only have access to $B_z$ meaning that $\nabla \times \mathbf{m}_{x,y}$, and in turn $C_z$, is fully undetermined.\\
Even if we had full knowledge of $\mathbf{F}$, a second degree of arbitrariness in the knowledge of the vector and scalar potential comes, as in EM, from the following \textit{gauge}-like degree of freedom:
\begin{align}
m_z(\bm{\uprho}) &= m'_z(\bm{\uprho}) + \Lambda, \nonumber \\
\mathbf{m}_{x,y}(\bm{\uprho}) &= \mathbf{m}'_{x,y}(\bm{\uprho}) - \alpha_{x,y}^{-1}(d,t) \ast \alpha_{z}(d,t) \ast \nabla \Lambda,
\label{eq:gauge}
\end{align}
where $\Lambda(\bm{\uprho},d)$ is an arbitrary function of space.\\
As explained in the main text, in order to fix the arbitrary functions $\Lambda(\bm{\uprho},d)$ and $C_z(\bm{\uprho},d)$ and therefore classify the different spin structures producing the measured stray field, we proceed in analogy with EM. Each physically distinct configuration of the spin texture is obtained after making local assumptions about the vector field $\mathbf{m}$, with a procedure that resembles standard gauge fixing in EM \cite{Griffith}.\\
Two of these possible assumptions, motivated by the spiral (cycloid) nature of Bloch (N\'eel) domain walls \cite{Tetienne15} and the resulting partial differential equations that need to be solved in order to determine $\mathbf{m}$ are reported in the next Sections. 
\subsection{Bloch or Coulomb effective gauge}
\label{sec:bloch}
By a solution in the Bloch gauge to the stray field equation, we mean a solution to \eqref{eq:gauss} for $\mathbf{m}$ in which we make the local assumption:
\begin{equation}
\nabla \cdot \mathbf{m}_{x,y} = 0,
\label{eq:BlochCondv2}
\end{equation}
whose physical justification has been given in the main text.\\
Since $\mathbf{m}_{x,y}$ plays the role of an effective vector potential, the condition \eqref{eq:BlochCondv2} reminds us of the Coulomb gauge in EM.
Exactly as in EM, in the Coulomb gauge the equation providing us with the scalar potential is the Poisson one:
\begin{equation}
- \frac{2 B_z(\bm{\uprho},d)}{\upmu_0 M_s} = \alpha_{z}(d,t) \ast \nabla^2 m_z(\bm{\uprho}),
\label{eq:PoiblochEW}
\end{equation}
easy to solve in Fourier space. We also know that the solution to \eqref{eq:PoiblochEW} is unique once boundary conditions are fixed \cite{Griffith}.  
Once $m_z(\bm{\uprho})$ is found, we can then obtain $\mathbf{m}_{x,y}$ by solving $\nabla \cdot \mathbf{m}_{x,y} = 0$. The complete partial non-linear differential equation in the azimuthal angle $\phi(\bm{\uprho})$ reads as:
\begin{align}
& \nabla \cdot \sqrt{m_s^2(\bm{\uprho}) - m_z^2(\bm{\uprho})}  \left( \begin{array}{c} \cos(\phi) \\ \sin(\phi) \end{array} \right) = \mathscr{B}(\phi,\bm{\uprho}) = 0.
\label{eq:BlochSolv2}
\end{align}
Eq. \eqref{eq:BlochSolv2} takes normalization to the space-dependent saturation magnetization $m_s(\bm{\uprho})$ into account (see also Section \ref{sec:calib} for more details on this last point). We obtain a solution to \eqref{eq:BlochSolv2} variationally, by minimizing the following cost function with respect to $\phi$:
\begin{align}
\mathscr{C}(\phi) = \int \mathscr{B}^2(\phi,\bm{\uprho}) \diff \bm{\uprho}.
\label{eq:CostBloch}
\end{align}
A very brief reminder of the popular steepest descent method we have used for minimizing the quadratic form in \eqref{eq:CostBloch} is presented in Section \ref{sec:steep}.
\subsection{N\'eel effective gauge}
\label{sec:neel}
By a solution in the N\'eel \textit{gauge} to the stray field equation, we mean a solution to \eqref{eq:gauss} for $\mathbf{m}$ in which we make the local assumption:
\begin{equation}
\nabla \times \mathbf{m}_{x,y} = 0.
\label{eq:NelCond}
\end{equation}
Fixing the curl of $\mathbf{m}_{x,y}$ is equivalent to fixing the curl for the effective electric field $\mathbf{F}$ or, equivalently, the function $C_z$ in \eqref{eq:effeF}. 
The vector field $\mathbf{F}$ becomes therefore conservative and it can be obtained explicitly from:
\begin{align}
& B_z(\bm{\uprho},d) = \nabla^2 V, \nonumber \\
& \mathbf{F} = - \nabla V. 
\label{eq:NelSolv}
\end{align}
At this point, an explicit solution to the stray field equation is still not possible as we retain the degree of freedom given by the arbitrary function $\Lambda(\bm{\uprho},d)$ (note that a transformation like the one in \eqref{eq:gauge} preserves the curl of the vector field $\mathbf{m}_{x,y}$). 
In order to further reduce the manifold of possible solutions, we introduce the normalization of the vector field $\mathbf{m}$ in the form of:
\begin{equation}
\mathbf{F}=  \frac{\upmu_0 M_s}{2} \left( \alpha_{z}(d,t) \ast \nabla m_z(\bm{\uprho})  + \alpha_{x,y}(d,t) \ast  \sqrt{m_s^2(\bm{\uprho}) - m_z^2(\bm{\uprho})} \mathbf{u}_{\phi} \right),
\label{eq:F2}
\end{equation}
with $\mathbf{u}_{\phi}$ the unit vector $(\cos(\phi),\sin(\phi))$. Eq. \eqref{eq:F2} represents two coupled non-linear partial differential equations in $\phi$ and $m_z$. In order to produce Fig.~3e of the main text we have solved it by minimizing (with respect to  $\phi$ and $m_z$) the following cost function, variationally:
\begin{align}
\mathscr{C}(\phi,m_z) = \int \left[ (F_x(\phi,m_z,\bm{\uprho}) + \partial_x V )^2 + (F_y(\phi,m_z,\bm{\uprho}) + \partial_y V)^2 \right] \diff \bm{\uprho}.
\label{eq:CostNeel}
\end{align}
In Section \ref{sec:unic} we discuss the degeneracy of the solution once normalization and curl have been fixed.

\section{Steepest descent minimization}
\label{sec:steep}
The minimization of a quadratic form using numerical, iterative steepest descent procedures is reported in several textbooks \cite{Martin04}, for instance in the context of energy functionals.\\
In general, it is well known that a quadratic function $\mathscr{C}(\{x_{\alpha}\})$ of $N$-variables $x_{\alpha}, \alpha=1 \ldots N$ can be minimized starting from the guess $x_{\alpha,0}$, by iteratively moving antiparallel to the gradient direction, e.g. \cite{Martin04}:
\begin{equation} 
x_{\alpha,i+1} = x_{\alpha,i} - \lambda \frac{\partial \mathscr{C}(\{x_{\alpha}\})}{\partial x_{\alpha}} \Big|_{x_{\alpha,i}}, 
\label{eq:steep}
\end{equation} 
where $\lambda$ is a constant and the $i=0, \cdots N_s$-index refers to the iteration number. The previous follows from the fact that gradients are orthogonal to isolines directions:
\begin{equation} 
\diff \mathscr{C}(\{x_{\alpha}\}) = 0 = \sum_{\alpha} \frac{\partial \mathscr{C}}{\partial x_{\alpha}} \diff x_{\alpha}.
\label{eq:steep2}
\end{equation} 
When $\mathscr{C}$ becomes a functional, $N \rightarrow \infty, \{x_{\alpha}\} \rightarrow \phi(\alpha)$ and the functional increment upon a change $\phi(\alpha)$ to $\phi(\alpha) + \eta(\alpha)$ can be written in first order as \cite{Binney}:
\begin{equation} 
\diff \mathscr{C}(\phi(\alpha)) = \int \diff \alpha  \frac{\delta \mathscr{C}}{\delta \phi(\alpha)} \eta(\alpha).
\label{eq:steep3}
\end{equation} 
The analogy between \eqref{eq:steep3} and \eqref{eq:steep2} allows to rewrite the update in \eqref{eq:steep} in the continuous limit as \cite{Martin04}:
\begin{equation} 
\phi_{i+1}(\alpha) = \phi_{i}(\alpha) - \lambda \frac{\delta \mathscr{C}}{\delta \phi(\alpha)}  \Big|_{\phi_{i}(\alpha)}, 
\label{eq:steep4}
\end{equation} 
where the derivative with respect to $\mathscr{C}$ is a functional one \cite{Binney}. The functionals that we have to minimize in this work have, like \eqref{eq:CostBloch} and \eqref{eq:CostNeel}, the general form:
\begin{equation} 
\mathscr{C}(\phi(\bm{\uprho})) = \int \mathscr{L}^2(\phi,\bm{\uprho})  \diff \bm{\uprho}.
\label{eq:steep5}
\end{equation} 
The functional derivative in \eqref{eq:steep4} can therefore be computed using chain derivatives as:
\begin{equation} 
\frac{\delta \mathscr{C}}{\delta \phi(\bm{\uprho})} = 2 \mathscr{L}(\phi,\bm{\uprho}) \frac{\delta \mathscr{L}}{\delta \phi(\bm{\uprho})}
 - 2 \nabla \mathscr{L}(\phi,\bm{\uprho}) \cdot \frac{\partial \mathscr{L}}{\partial \nabla \phi(\bm{\uprho})}.
\label{eq:steep6}
\end{equation} 
We then numerically implement \eqref{eq:steep4} in order to obtain the function $\phi$ that minimizes $\mathscr{C}$.
\section{Uniqueness of the solution at fixed gauge}
\label{sec:unic}
%
Fixing the curl of $\mathbf{m}_{x,y}$ and locally imposing a normalization of the ordered moment does not guarantee that the solution reproducing a given target stray field will be unique.\\
In this Section we first discuss which kind of transformations would preserve the curl and normalization of the vector field and will finally briefly comment on the uniqueness of the solution.
\subsection{General Transformation preserving the curl}
As known from standard EM, a gauge transformation like \eqref{eq:gauge} would preserve the curl of the magnetic structure and the field it produces. On the other hand, in general that same transformation does not preserve the normalization of the vector field. We start rewriting \eqref{eq:gauge} in the following form:
\begin{align}
m_{1,z} &= m_{2,z} + \Lambda, \nonumber \\
\mathbf{m}_{1,x,y} &= \mathbf{m}_{2,x,y} + \nabla \Lambda'. 
\label{eq:gauge2}
\end{align}
If we want to locally have $||\mathbf{m}_{1}|| = ||\mathbf{m}_{2}||$ we find that the following must hold: 
\begin{equation}
2 \mathbf{m}_1 \cdot \left( \begin{array}{c} \nabla \Lambda' \\  \Lambda \end{array} \right) = \left( \begin{array}{c} \nabla \Lambda' \\  \Lambda \end{array} \right) \cdot \left( \begin{array}{c} \nabla \Lambda' \\  \Lambda \end{array} \right).
\label{eq:norm5}
\end{equation}
The previous equation fixes a condition for the norm of the vector $\Delta \mathbf{m} = (\nabla \Lambda',\Lambda)$. In particular, one can see that if we assume $||\mathbf{m}_{1}|| = ||\mathbf{m}_{2}||=1$, then $|| \Delta \mathbf{m} || = 2 \cos(\theta_{1} - \theta_{\Delta \mathbf{m}})$, where $\theta_{1} - \theta_{\Delta \mathbf{m}}$ is the angle between the $\mathbf{m}_1$ and the $\Delta \mathbf{m}$ vector.
\subsection{Special cases}
Eq. \eqref{eq:norm5} can be easily solved for $\Lambda$ in special cases, which allow us to prove that in general fixing the curl of $\mathbf{m}_{x,y}$ and locally imposing a normalization of the ordered moment does not automatically guarantee a unique solution for the magnetic pattern.\\ 
Assume for instance that $\theta_{1} = 0$ everywhere in space, meaning we are considering a ferromagnetic pattern, which clearly has zero curl. Eq. \eqref{eq:norm5} then reduces to:
\begin{equation}
\alpha_{z} \ast \nabla \left( 2 \cos^2(\theta_{\Delta \mathbf{m}}) \right) = - \alpha_{x,y} \ast \left( \sin(2 \theta_{\Delta \mathbf{m}}) \mathbf{u}_{\phi_{\Delta \mathbf{m}}} \right),
\label{eq:condPF}
\end{equation}
where $\mathbf{u}_{\phi} = (\cos(\phi_{\Delta \mathbf{m}}),\sin(\phi_{\Delta \mathbf{m}}))$ contains the azimuthal angle of the $\Delta \mathbf{m}$ vector.
Finding a solution to \eqref{eq:condPF} is complicated by the presence of convolutions. \\
We make the only assumption that the sinusoidal functions in $\theta_{\Delta \mathbf{m}}$ have a Fourier spectrum centred around $\mathbf{\tilde{k}}$, such that the convolutions can be approximated with multiplications:
\begin{equation}
\alpha_{z}(\mathbf{\tilde{k}}) \nabla \left( \cos(2  \theta_{\Delta \mathbf{m}}) \right) = - \alpha_{x,y}(\mathbf{\tilde{k}}) \left( \sin(2 \theta_{\Delta \mathbf{m}} ) \mathbf{u}_{\phi} \right).
\label{eq:condPF3}
\end{equation}
We can now evaluate the ratio $\alpha_{x,y}(\mathbf{\tilde{k}})/\alpha_{z}(\mathbf{\tilde{k}})$ using the results in Section \ref{sec:alphas} and obtain:
\begin{equation}
\nabla \theta_{\Delta \mathbf{m}} = \frac{\tilde{k}}{2} \mathbf{u}_{\phi_{\Delta \mathbf{m}}}.
\label{eq:condPF4}
\end{equation}
So, in essence, good solutions for $\Delta \mathbf{m}$ are those in which its polar angle has a gradient with a constant norm. Functions linear with the spatial coordinate will be solutions, such as plane waves $\theta_{\Delta \mathbf{m}} = (\mathbf{\tilde{k}} \cdot \mathbf{r})/2$ with $\mathbf{u}_{\phi_{\Delta \mathbf{m}}} \parallel \mathbf{\tilde{k}}$ or radial waves $\theta_{\Delta \mathbf{m}} = \tilde{k} r/2$. All these solutions are therefore cycloids, in the definition given in the main text. \\
We found that cycloids with a single wave vector and constant ordered moment can actually produce no stray field, as ferromagnetic states also do not according to \eqref{eq:Bz}. \\
On the other hand, we also note that if we fix boundary conditions, e.g. assume the magnetic moments to be in a ferromagnetic state for $r \rightarrow \infty$, then in order to avoid having $B_z \neq 0$ at the boundary between the ferromagnetic and the plane wave region we need to select $\tilde{k} \rightarrow 0$. \\
We therefore argue that fixing the curl of $\mathbf{m}_{x,y}$, then locally imposing a normalization and boundary conditions selects a unique solution to the stray field equation.

\section{Continuous tuning of the magnetic structure}
The solutions presented in Section \ref{sec:bloch} and \ref{sec:neel} are only two special cases of the infinitely many $\mathbf{m}$ satisfying \eqref{eq:Bz} given a stray field $B_z$. In order to continuously explore the solution manifold, we start from the Bloch $\mathbf{m}_{\mathrm{B}}$ and first perform a local rotation $\mathbf{\bar{R}}_z(\Delta \phi(\lambda)) \mathbf{m}_{\mathrm{B}}$  of the magnetic structure about the $z$-axis by an angle $\Delta \phi(\lambda)$, defined as: 
\begin{equation}
\Delta \phi(\lambda) = \phi_{\mathrm{B}} + \lambda \left( \phi_{\mathrm{N}}  - \phi_{\mathrm{B}} \right).
\label{eq:df}
\end{equation}
In \eqref{eq:df}, $\phi_{\mathrm{B}}, \phi_{\mathrm{N}}$ are the solutions for the azimuthal angles in \eqref{eq:CostBloch} and \eqref{eq:CostNeel} for the Bloch and N\'eel case, respectively, and $0 \leq \lambda \leq 1$ is a constant. As $\lambda \rightarrow 1$, the orientation of the in-plane local moments will be parallel to the one obtained in the N\'eel solution; however, the stray field produced by the resulting spin structure $\mathbf{m}_{\lambda, \mathrm{B}} = \mathbf{\bar{R}}_z(\Delta \phi(\lambda)) \mathbf{m}_{\mathrm{B}}$ will not match the target field measured in experiments.\\
In order to preserve the in-plane orientation and match the target field, starting from $\mathbf{m}_{\lambda, \mathrm{B}}$ we locally rotate the structure about the locally varying in-plane axis $\mathbf{u}(\bm{\uprho})$ perpendicular to $\mathbf{m}_{\lambda, \mathrm{B}}$ and defined as:
\begin{equation}
\mathbf{u} = \frac{1}{|| \mathbf{m}^{x,y}_{\lambda, \mathrm{B}} ||} \left( \begin{array}{c} m^y_{\lambda, \mathrm{B}} \\ - m^x_{\lambda, \mathrm{B}} \\  0 \end{array} \right).
\label{eq:rotvar2}
\end{equation}
Rotations $\mathbf{\bar{R}}_{\mathbf{u}}$ by an angle $\theta(\bm{\uprho})$ about the local axis $\mathbf{u} \perp \mathbf{m}_{\lambda, \mathrm{B}}$ can be readily expressed using the Rodriguez's formula \cite{Rodbook}:
\begin{equation}
\mathbf{\bar{R}}_{\mathbf{u}} = \cos(\theta) \mathbf{\bar{I}} + \sin(\theta) [\mathbf{u}]_{\times},
\label{eq:rod}
\end{equation}
where $[\ldots]_{\times}$ is the cross product matrix and $\mathbf{\bar{I}}$ is the identity. After the $\mathbf{\bar{R}}_z(\Delta \phi(\lambda))$ and $\mathbf{\bar{R}}_{\mathbf{u}}$ rotation, the final structure reads as:
\begin{equation}
\mathbf{m}_{\theta, \lambda, \mathrm{B}} = \mathbf{\bar{R}}_{\mathbf{u}} \mathbf{\bar{R}}_z(\Delta \phi(\lambda)) \mathbf{m}_{\mathrm{B}} = \cos(\theta) \mathbf{m}_{\lambda, \mathrm{B}} + \sin(\theta) \mathbf{p}_{\lambda, \mathrm{B}}, 
\label{eq:rod2}
\end{equation}
where the vector $\mathbf{p}_{\lambda, \mathrm{B}}$ is a vector orthogonal to and with the same norm of $\mathbf{m}_{\lambda, \mathrm{B}}$, that can be written as:
\begin{equation}
\mathbf{p}_{\lambda, \mathrm{B}} =  \left( \begin{array}{c}  - \frac{m^z_{\lambda, \mathrm{B}} m^x_{\lambda, \mathrm{B}}}{|| \mathbf{m}^{x,y}_{\lambda, \mathrm{B}}  ||} \\ -  \frac{m^z_{\lambda, \mathrm{B}} m^y_{\lambda, \mathrm{B}}}{|| \mathbf{m}^{x,y}_{\lambda, \mathrm{B}} ||} \\ || \mathbf{m}^{x,y}_{\lambda, \mathrm{B}} || \end{array} \right).
\label{eq:rotvar3}
\end{equation}
%
%
%
If now we assume that a certain magnetic structure $\mathbf{m}$ produces a stray field $B_z(\mathbf{m})$, then a solution for $\theta(x,y)$ in \eqref{eq:rod} can be obtained by minimizing with a numerical variational analysis (see Section \ref{sec:steep}) the following functional:
\begin{equation} 
\mathscr{C}(\theta) = \int \left( B_z(\mathbf{m}_{\theta, \lambda, \mathrm{B}})  - B_z(\mathbf{m}_{\mathrm{B}} )  \right)^2 \diff \bm{\uprho} .
\end{equation} 
The resulting function $\theta$ is found to be continuous and it's used to compute the solutions in Fig.~4 of the main text.
\section{Calibrations}
\label{sec:calib}
\begin{figure}[h!]
\centering
\includegraphics[width=1\textwidth]{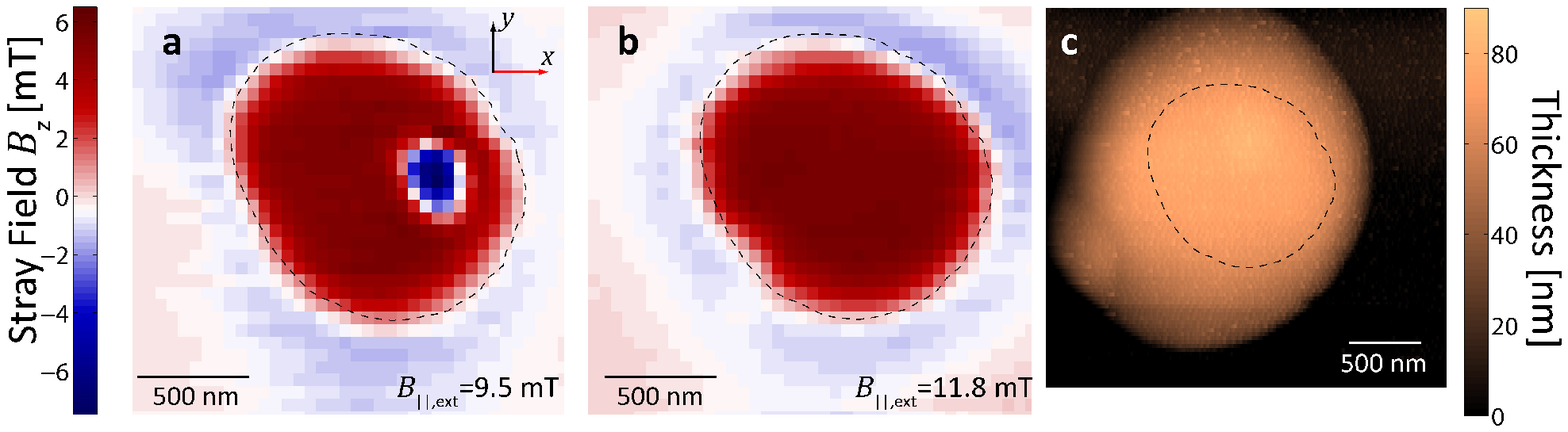}
\caption{\textbf{Comparison between the region with a magnetic signal and topography.} \textbf{a:} Reconstructed $B_z$ stray field measured with a 9.5 mT bias field (see main text for details). The black dashed line outlines the boundary of the region within which the largest stray field is recorded \textbf{b:} Same as in \textbf{a}, but with a bias field of 11.8\ mT. The black dashed line is the same as in \textbf{a}. We clearly see that the bubble-like feature has disappeared at this field. Note also that the stray field $B_z$ is qualitatively constant while moving along the boundary of the magnetic disc, supporting the assumptions that at these fields the magnetization is mostly out-of-plane. \textbf{c:} Surface topography recorded by monitoring the piezo voltage $V_z$. The dashed black boundary is the same as in \textbf{a} and \textbf{b}, stressing the fact that the region in which we observe a magnetic stray field is smaller than the actual physical disc size.}
\label{fig:Bound}
\end{figure}
\begin{figure}[h!]
\centering
\includegraphics[width=0.8\textwidth]{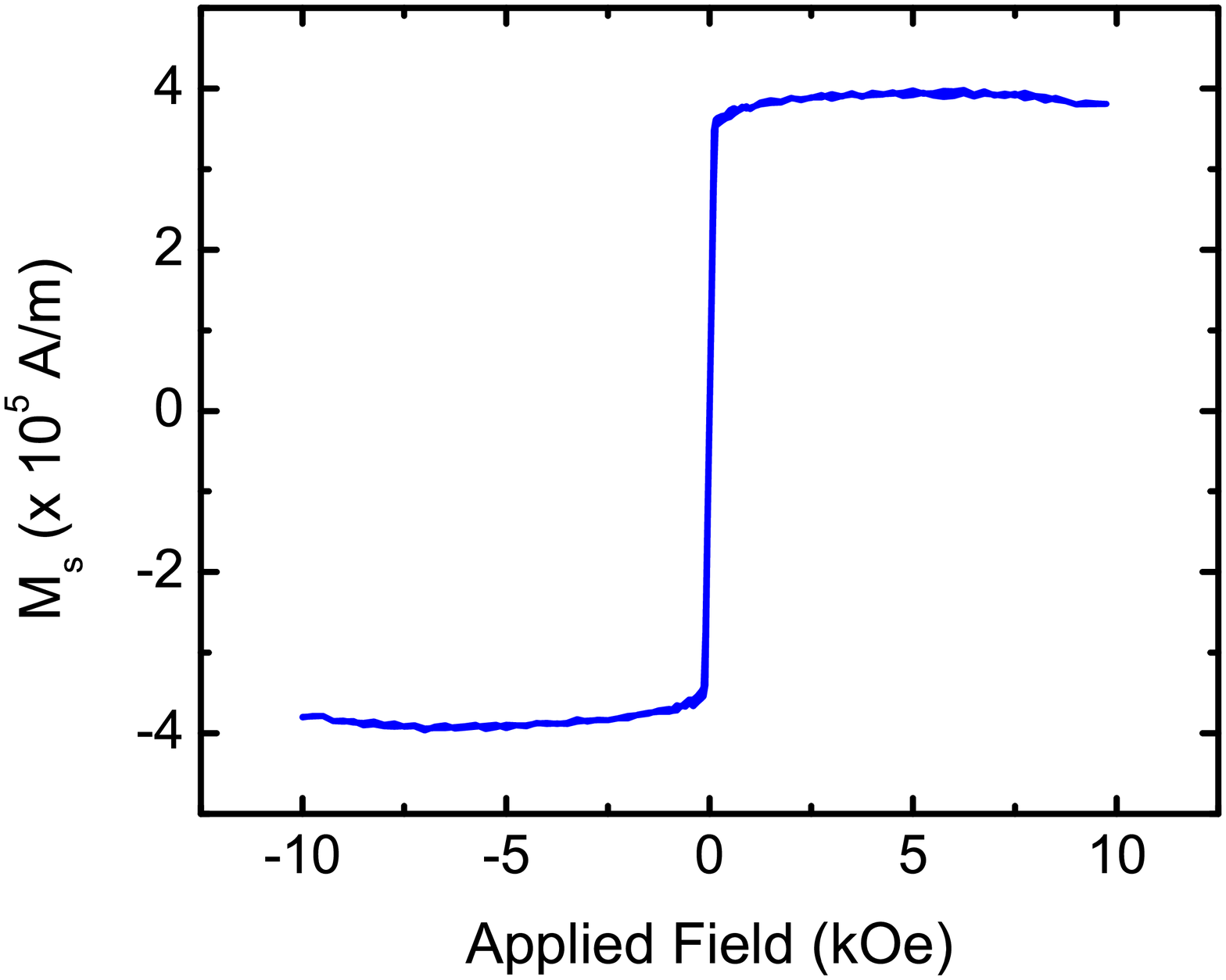}
\caption{\textbf{Field-dependent magnetization of the reference Si substrate.} Magnetization measured via Vibrating Sample Magnetometry (VSM) of a Si substrate held at the same height as the surface of the quartz tip during sputtering. The measurement reveals a bulk magnetization at saturation of the order of $M_s \simeq 3.8 \cdot 10^5$ A/m.}
\label{fig:MsVSM}
\end{figure}
In our work we have carried out a reconstruction of the underlying local magnetization configuration starting from eq. \eqref{eq:Bz}. The unknown parameters in the equation are the film thickness $t$, the NV depth $d$ and the local value of the saturation magnetization $M_s \cdot m_s(\bm{\uprho})$, which has been used for the reconstruction in eq. \eqref{eq:BlochSolv2} and \eqref{eq:F2}.\\
In order to calibrate these values we start from a simultaneous measurement of the magnetic disc topography and stray field map at saturation, as shown in Fig.~\ref{fig:Bound}. We first compare the stray field maps in Fig.~\ref{fig:Bound}a and Fig.~\ref{fig:Bound}b with the surface topography measured by monitoring the vertical movement of the tip, shown in Fig.~\ref{fig:Bound}c. In each image we superimpose a black dashed boundary qualitatively representing the region within which a magnetic signal is measured. By comparison of this boundary with the surface topography, we see that magnetic signal is measured from the region in the disc having a constant thickenss. We conclude that within the field of view in Fig.~\ref{fig:Bound}a and Fig.~\ref{fig:Bound}b, it is the saturation magnetization $M_s \cdot m_s(\bm{\uprho})$ that varies and not the film thickness $t$. We therefore retain $t$ as constant in eq. \eqref{eq:Bz} and make use of eq. \eqref{eq:res5} in order to compute the resolution functions. In particular, the values used during the deposition are $t = 1.1$~nm, $N=10$, $s = 7$~nm, in agreement with the measured total thickness of the film in Fig.~\ref{fig:Bound}c. Note that for the NV depth $d$ we use $d \sim$ 30~nm, a value that SRIM calculations predict to be in agreement with the 18 keV implantation energy of our diamond \cite{Spinicelli11}.\\   
We now assume the magnetization to be out-of-plane due to magnetic anisotropy\cite{woo16} in the regime in which the skyrmion disappears; such assumption is well supported by looking at the spatially homogeneous stray field pattern for $B_z$ measured at the magnetic disc edge in Fig.~\ref{fig:Bound}b. 
With this information we can now estimate the local value of the saturation magnetization for Co. By a direct inversion of the $B_z$ profiles in Fig.~\ref{fig:Bound}a and Fig.~\ref{fig:Bound}b, i.e. solving eq. \eqref{eq:PoiblochEW} for $M_s \cdot m_z$ in both regimes, we obtain Fig.~3d of the main text. It should be pointed out that in order to obtain the inversion at saturation we solve eq. \eqref{eq:PoiblochEW} and work in the Bloch gauge because in this regime $\mathbf{m}_{x,y}=0$, which therefore satisfies $\nabla \cdot \mathbf{m}_{x,y}=0$.
The $M_s \cdot m_z(\bm{\uprho})$ value at 11.8 mT (saturation) is equivalent to $M_s \cdot m_s(\bm{\uprho})$, telling us how is the nominal $M_s$ value locally renormalized ($ m_s(\bm{\uprho})$) due to variations in the saturation magnetization. With this procedure we estimate (see Fig.~3d of the main text) a maximum value for $M_s \cdot m_s(\bm{\uprho})$ of $M_s \simeq 3.6 \cdot 10^5$ A/m at the disc centre. 
We then independently measured, using Vibrating Sample Magnetometry (VSM), the nominal value of $M_s$ for our film by means of a reference Si wafer placed in the sputtering chamber together with our tip during the deposition process. We found $M_s \simeq 3.8 \cdot 10^5$ A/m (see Fig.~\ref{fig:MsVSM}), in agreement with the NV measurement. Note that such value for $M_s$ is less than half the bulk value, suggesting a magnetic dead layer due to roughness or oxidation. In-plane (hard axis) measurements (not shown) revealed saturation fields of $\approx 5$ kOe.

\section{Stability of the solution: NV depth}
\label{sec:Depth}
\begin{figure}[h!]
\centering
\includegraphics[width=1\textwidth]{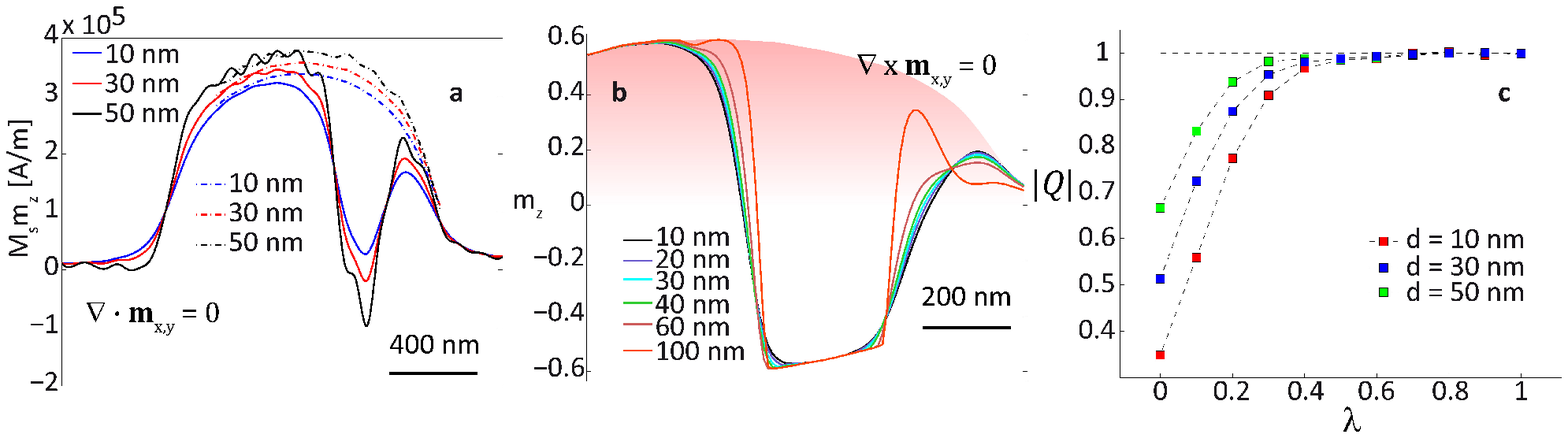}
\caption[itemsep=-1mm]{\textbf{Stability of the solution upon varying the NV depth.} \textbf{a:} Cut through the reconstructed out-of-plane magnetization in the Bloch gauge. The solid (dashed) lines represent solutions at the two different bias fields of 9.5 (11.8) mT. As the NV depth increases the noise in m$_z$ also increases due to the exponential prefactor in eq. \eqref{eq:dipoDefKernel}. \textbf{b:} Cut through the skyrmion core displaying the reconstructed magnetic structure in the N\'eel gauge for different NV depths $d$ (solid lines). As $d$ increases the m$_z$ profile near the skyrmion edges gets steeper in order to keep the resulting stray field constant. \textbf{c:} Absolute value of the topological number for different NV depths, while tuning the spin configuration from Bloch to N\'eel with the parameter $\lambda$ defined in eq. \eqref{eq:df}.}
\label{fig:NVdepth}
\end{figure}
Based on the implantation energy of the diamond used in our experiments, NV centres in our pillars are expected to be present at an estimated depth of $d \sim$30 nm below the surface. The aim of this section is to study the stability of the reconstructed solutions upon a change of the NV-sample distance. Intuitively, as the stray field $B_z$ is directly related to the gradient of the local magnetization, for increasing values of $d$ the spatial variations of the local magnetization will have to increase in order for $B_z$ to remain constant. This is what we observe, e.g., in Fig.~\ref{fig:NVdepth}a, where solutions at different values of $d$ are obtained in the Bloch gauge for two values of the applied field. As one can see, the solutions at larger values of $d$ in the Bloch gauge (obtained by a direct solution of a Poisson-like equation in Fourier space) are affected by more noise at high wavenumbers due to the problem of \textit{downward propagation}\cite{Blakely96}. In addition, we notice that at saturation the magnetization varies by only $\approx \pm 6$\% for an NV depth change of $\pm 20$ nm. \\
In order to avoid the forward propagation issue and check for the stability of the solution obtained in the N\'eel gauge upon a change in the NV depth, we approximate $M_s \simeq 3.6 \cdot 10^5$ A/m (as shown in Fig.~3d of the main text and consistently with the magnetometry data in Fig.~\ref{fig:MsVSM}) and obtain the set of solutions in Fig.~\ref{fig:NVdepth}b. Qualitatively, we can see the skyrmion walls getting slightly sharper with larger NV distance.
Regardless, we always obtain a domain-wall like solution for the skyrmion, with the same characteristic diameter $\rho_0 \approx 200$ nm. In Fig.~\ref{fig:NVdepth}c we plot the topological number as a function of the parameter $\lambda$ defined in eq. \eqref{eq:df} for different values of the parameter $d$. This analysis of the topology of the solution allows us to isolate the N\'eel configuration even when the parameter $d$ is modified. 

\section{Micromagnetic simulations: stability of the N\'eel caps}
\label{sec:Ncap}
\begin{figure}[h!]
\centering
\includegraphics[width=0.7\textwidth]{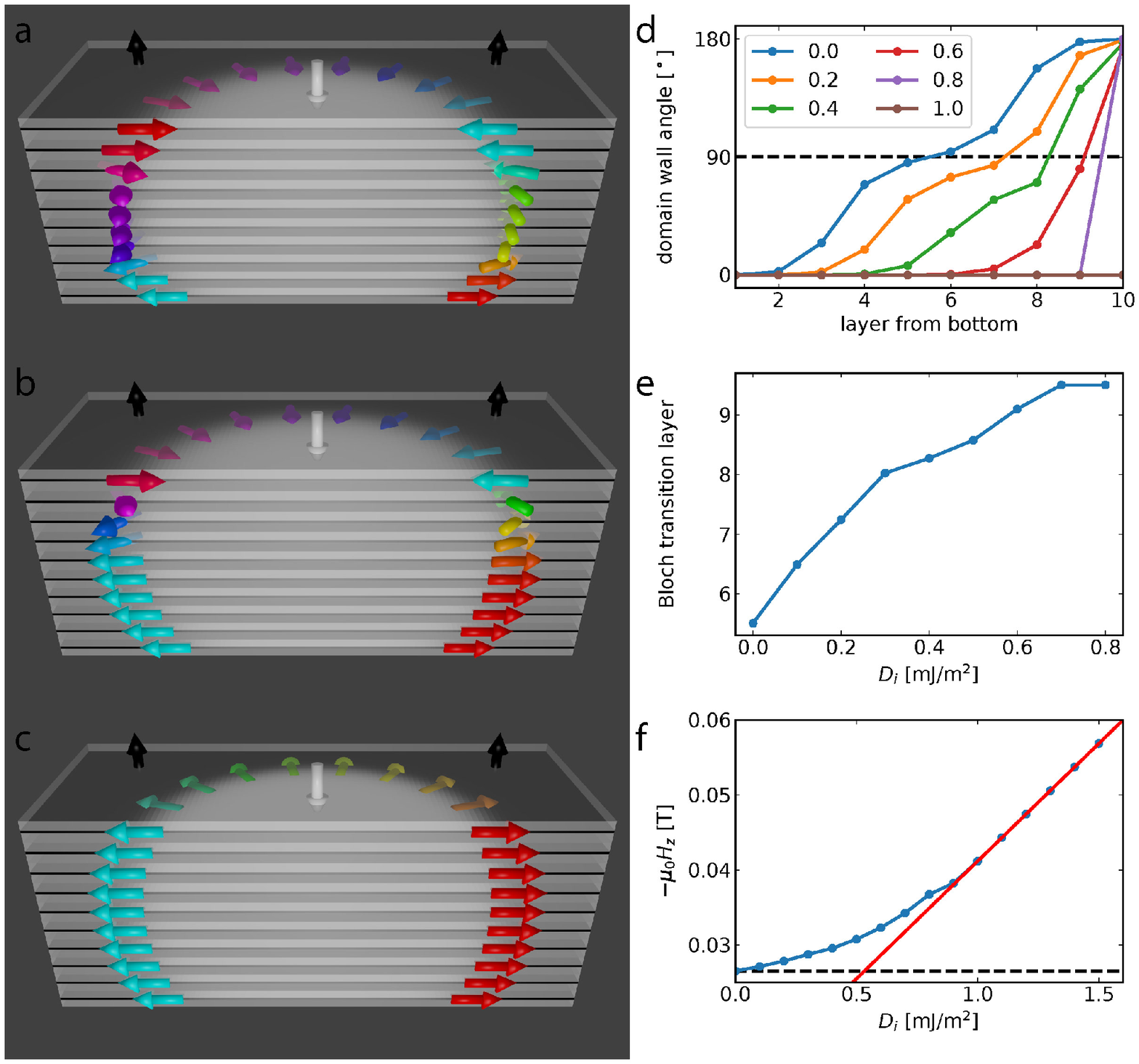}
\caption[itemsep=-1mm]{\textbf{Helicity change along $z$ in multilayers with different DMI.} \textbf{a-c}: 3D cross section through the center of a skyrmion spin structure in the same multilayer as in Fig. 5 of the main text, with DMI values of $D_i=0$ (\textbf{a}), $D_i=0.5$ mJ/m$^2$ (\textbf{b}), and $D_i={1.0}$ mJ/m$^2$ (\textbf{c}). The colored arrows indicate the magnetization orientation in and position of the domain wall surrounding the skyrmion in each respective layer. The skyrmion radius is largest in the central layers. \textbf{d}: Domain wall angle (here equivalent to the helicity) for the individual layers as a function of DMI, as indicated by the line color. The legend specifies the DMI value for each color in units of mJ/m$^2$. \textbf{e}: Layer in which the domain wall angle is 90$^{\circ}$, where fractional values are obtained by linear interpolation of the data in \textbf{d}. \textbf{f}: Field required to stabilize a $R=50$ nm radius skyrmion in the multilayer as a function of DMI. For large DMI, i.e., once all layers have a left-handed chirality, the field scales linearly with $D_i$, as indicated by the red line. For $D_i<1$ mJ/m$^2$, the required field is larger than expected from a material with uniform magnetization along $z$.}
\label{fig:S6}
\end{figure}
To understand the origin of the observed right-handed chirality in a material with left-handed DMI, we performed full 3D micromagnetic simulations of a skyrmion in a multilayer system with ten magnetic layers. Here, we assume a CoFeB-based system with a saturation magnetization of $M_s=10^6$ A/m, an exchange constant of $A=10$ pJ/m, an anisotropy field of $\mu_0H_k=0.2$ T and a variable strength of the DMI $D_i$. Each of the ten repeats consists of 1 nm of magnetic material and 7 nm of non-magnetic spacers, such as Pt and Ta. Each repeat is simulated by a single cell in $z$ direction by using the effective medium model \cite{woo16,Lemesh17}. We set a tiny inter-layer exchange couping of 0.1 pJ/m to break the degeneracy between clockwise and counterclockwise Bloch configurations. Generally, the parameters were chosen to obtain skyrmions with 50 nm radius in an out-of-plane magentic field of about 50 mT.

Fig.~\ref{fig:S6} shows the final magnetic state at different values of DMI after relaxing a skyrmion state similar to Fig.~\ref{fig:S6}a for at least 20 ns in the magnetic field shown in Fig.~\ref{fig:S6}f. For DMI values $D_i<1$ mJ/m$^2$ (Fig.~\ref{fig:S6}c) we observe a right-handed chirality in the top-most layer. The domain wall angle in each layer as a function of $D_i$ is plotted in Fig.~\ref{fig:S6}d. A domain wall angle $>90^{\circ}$ indicates a right-handed wall. The lower the DMI the more layers show a right-handed chirality. The layer number in which the chiralty switches from left-handed to right-handed (i.e., the layer in which the configuration is purely Bloch-like) is plotted in Fig.~\ref{fig:S6}e as a function of $D_i$. It is important to note that, in contrast to other surface-sensitive techniques, such as Photo-Emission Electron Microscropy (PEEM), NV magnetometry can distinguish beween the various cases shown in Fig.~\ref{fig:S6}d even though the top layer magnetization remains the same. 

For a uniform magnetization along $z$, the field required to stabilize a skyrmion of a given size ($R=50$ nm in the present case) is proportional to $D_i$. This behavior is confirmed in the high DMI regime in Fig.~\ref{fig:S6}f, i.e., for DMI values where all layers have the same chirality. The formation of flux-closure domains, however, breaks this trend. Specifically, the stabilizing field of $26$ mT for the zero DMI case could be misinterpreted as a DMI strength of $D_i\approx 0.5$ mJ/m$^2$ if a uniform magnetization along $z$ is assumed (see red line in Fig.~\ref{fig:S6}f). This observation underlines the significance of volume stray field interactions and flux closure domains for the interpretation and design of skyrmions in magnetic multilayers.

\section{Global versus local effective gauge fixing}
\label{sec:localgauge}
In section~\ref{sec:gaugeFix} an effective gauge was imposed in order to find a solution for the local magnetization. Conditions such as those in eq. \eqref{eq:BlochCondv2} and \eqref{eq:NelCond} were imposed \textit{globally} through the magnetic stack, meaning that the magnetization pattern $\mathbf{m}(\bm{\uprho})$, the pattern's helicity and chirality were the same regardless the magnetic layer considered.\\
In the main text we compare our experimental data with a model in which the top and bottom three layers have opposite N\'eel chirality, contrary to the Bloch-like intermediate layers. This case can be easily considered starting from eq. \eqref{eq:res5} in section \ref{sec:alphas}, introducing parameters $c_i$ such that now: 
\begin{align}
& \alpha_{x,y}(d,t) \rightarrow \alpha_{x,y,N}(d,t) = \sum_{\nu =0}^{N-1} c_{\nu} \alpha_{x,y}(d + \nu\cdot s,t). 
\label{eq:res5VarChi}
\end{align}
Opposite chirality between, say, layer $i$ and $j$ can be imposed by simply setting $c_i=-c_j$. This results in a new $\alpha_{x,y}(d,t)$, to which we apply the formalism discussed in section~\ref{sec:gaugeFix}. In this way, the magnetization pattern $\mathbf{m}(\bm{\uprho})$ can still be considered as layer-independent but the in-plane magnetization will be summed up oppositely between the layers $i$ and $j$; within the minimization process, this effectively inverts the relative chirality between the layers. In order to account for the intermediate layers hosting Bloch-like skyrmions, we have decided to run the minimization process still within the N\'eel gauge, but setting for the intermediate $i$-layer the coefficients $c_i=0$. This condition implies that the term $\nabla \cdot \mathbf{m}_{x,y}$ will not contribute, for those layers, to $B_z$, as it should be for a real Bloch solution. The last method allows us to obtain a layer-independent $m_z$ profile, which we plot in Fig.~5c of the main text.

\end{document}